\documentclass[%
 reprint,
 amsmath,
 amssymb,
 aps,
]{revtex4-1}
\usepackage{graphicx}% Include figure files
\usepackage{dcolumn}% Align table columns on decimal point
\usepackage{bm}% bold math
\usepackage{here}
\usepackage[large, FIGTOPCAP]{subfigure}
\usepackage{color}

\begin{document}

\preprint{APS/123-QED}

\title{Dynamical-systems theory of cellular reprogramming}% Force line breaks with \\

\author{Yuuki Matsushita}
\author{Tetsuhiro S. Hatakeyama}%
\author{Kunihiko Kaneko}
 \email{kaneko@complex.c.u-tokyo.ac.jp}
\affiliation{%
Department of Basic Science, Graduate School of Arts and Sciences, University of Tokyo, 3-8-1 Komaba, Meguro-ku, Tokyo 153-8902, Japan
%  Authors' institution and/or address\\
%  This line break forced with \textbackslash\textbackslash
}%

\date{\today}% It is always \today, today,

\begin{abstract}
In cellular reprogramming, almost all epigenetic memories of differentiated cells  are erased by the overexpression of few genes, regaining pluripotency, potentiality for differentiation.
Considering the interplay between oscillatory gene expression and slower epigenetic modifications, such reprogramming is perceived as an unintuitive, global attraction to the unstable manifold of a saddle, which represents pluripotency. The universality of this scheme is confirmed by the repressilator model, and by gene regulatory networks randomly generated and those extracted from embryonic stem cells.
\end{abstract}

% \keywords{Suggested keywords}

\maketitle

In the development of multicellular organisms, cells with identical genomes differentiate into distinct cell types.
This cellular differentiation process has often been explained as balls falling down the epigenetic landscape, as originally proposed by Waddington \cite{waddington1957strategy}:
balls start from the top of the landscape, and as development progresses, they fall into distinct valleys, which correspond to differentiated cell types.
In modern biology, such landscapes are believed to be formed by epigenetic regulation, including DNA and chromatin modifications\cite{bird2007perceptions, goldberg2007epigenetics, cortini2016physics, surani2007genetic}.
For pluripotent cells, these modifications are small, whereas each differentiated cell type has a different epigenetic modification pattern \cite{meshorer2006hyperdynamic, atlasi2017interplay, mikkelsen2007genome, tripathi2019chromatin}.
Cells with pluripotency, such as embryonic stem (ES) cells, are located in the vicinity of the first branching point into the valleys, because they can easily differentiate into different types of cells with just slight stimuli \cite{levenberg2002endothelial}.

In 2006, a seminal study by Takahashi and Yamanaka reported that differentiated cells can regain pluripotency only by overexpressing few genes (so-called the four {\it Yamanaka factors}) without direct manipulations of epigenetic modifications.
This was termed as reprogramming of induced pluripotent stem (iPS) cells \cite{takahashi2006induction}.
The reprogramming is often described as ``climbing'' the epigenetic landscape \cite{takahashi2016decade, 10.1242/dev.020867, velychko2019excluding}.
This hypothesis, however, has two problems that still need to be addressed:
(1) cells have many degrees of freedom, with expression and epigenetic modifications of many genes, whereas reprogramming manipulation involves only few degrees of freedom.
How is it possible?; moreover, 
(2) if the initial pluripotent state is represented by the top of the landscape, it is not a stable point.
Thus, how can reprogramming robustly make the cells head toward such an ``unstable'' state?

Theoretically, these issues should be resolved based on dynamical systems theory.
The interplay between fast gene regulation and slow epigenetic dynamics shapes the epigenetic landscape, and differentiated cells are represented by different attractors \cite{kauffman1969metabolic, wang2010potential, forgacs2005biological, forgacs2005biological}.
Therefore, upon the reprogramming operation, cellular states starting from different attractors first converge into a unique pluripotent state, which is not stable, from which states move toward various attractors afterwards.
At first glance, these requirements seem to be incompatible; an unstable state (e.g., repeller) is not attracted from different initial conditions. Hence, to satisfy these requirements, the pluripotent cell is expected to be represented at least by a saddle that is attracted from many directions and departs only along unstable directions (manifolds), which
represent the cell differentiation process, leading to attractors of different destinations.
To regain pluripotency by reprogramming, cellular states must be placed on the stable manifold of the saddle by common manipulations from different attractors.
Such manipulation, however, would require fine-tuned control.
In contrast, reprogramming is mediated by the overexpression of a few common genes across a variety of differentiated cell types.
Therefore, some dynamical systems concept beyond just a saddle is needed.

A recent experiment provides some clues on this subject. 
Temporal oscillations in DNA methylation and corresponding gene expression levels are observed during cellular differentiation \cite{rulands2018genome, kangaspeska2008transient}.
In fact, gene expression oscillations have also been reported during somitogenesis and in embryonic stem cells \cite{palmeirim1997avian, kobayashi2009cyclic, canham2010functional}, whereas its relevance to cell differentiation has been theoretically investigated for decades \cite{furusawa2012dynamical, ullner2008multistability, koseska2010cooperative, suzuki2011oscillatory, koseska2013transition, goto2013minimal, miyamoto2015pluripotency}.
Recalling possible significance of oscillatory dynamics, it is reasonable to consider that if there is an oscillation of fast gene expression around the saddle point of the slow epigenetic dynamics, global attraction to it from broad initial conditions may be attained beyond its stable manifold.
As the oscillation dynamics are extended beyond the stable manifold of a saddle, global attraction to the vicinity of the saddle may be possible by taking advantage of the interplay between fast expression and slow modification dynamics.

Herein, we verified this possibility by using a dynamical system model with a gene regulatory network (GRN) and epigenetic modification.
We consider a cell model in which the cellular state was represented by the expression $x_i$ and epigenetic modification level $\theta_i$ for each gene $i$, with $i = 1, 2, \dots, N$.
Gene expression dynamics, with faster time scales, are governed by GRN with mutual activation or inhibition by transcription factors \cite{mjolsness1991connectionist, salazar2000gene,salazar2001phenotypic, huang2005cell, kaneko2007evolution}, whereas slower epigenetic dynamics change the feasibility of gene expression, which follows the gene expression patterns.
We assumed the epigenetic feedback reinforement, meaning that as more a gene is expressed (silences), the more feasible (harder) to express. This hypothesis was based on the experimental observations on the Trithorax (TrxG) and Polycomb (PcG) group proteins, two of the essential epigenetic factors for cellular differentiation \cite{angel2011polycomb}.
Specifically, we adopted:
\begin{subequations}
\begin{align}
\frac{d x_i}{d t} &= F\left( \sum_j J_{ij} x_j + \theta_i + I_i(t) \right) - x_i  \label{dx}, \\
\frac{d \theta_i}{d t} &= \frac{1}{\tau} (x_i - \theta_i) \label{dtheta}.
\end{align}
\end{subequations}

In Eq. (\ref{dx}), gene expression shows an on-off response to the input by adopting the function $F(z) = \tanh (\beta z)$, whereas $\beta = 40$
\footnote{Although we adopted a symmetric function, the result to be discussed is not changed if asymmetry functions, including the Hill function, are introduced.}.
If $J_{ij}$ is positive (negative), gene $j$ activates (inhibits) gene $i$, whereas $J_{ij}$ is set to $0$ if no regulation exists.
External input $I_i(t)$ is applied only during reprogramming manipulation to flip the expression of the gene $i$.
For simplicity, $I_i(t)$ takes a constant non-zero value when gene $i$ is overexpressed for reprogramming manipulation and zero otherwise.

In Eq. (\ref{dx}), $-\theta_i$ works as the threshold of the expression of the gene $i$, which represents the epigenetic modification status (when there is no epigenetic modification, it takes zero).
Eq. (\ref{dtheta}) represents the epigenetic feedback regulation.
Following the experimental observation of positive epigenetic feedback \cite{hihara2012local, grunstein1998yeast, schreiber2002signaling, dodd2007theoretical, sneppen2008ultrasensitive, angel2011polycomb, spainhour2019correlation}, we adopted this simple form as its specific form has not yet been confirmed \cite{furusawa2013epigenetic, gombar2014epigenetics, miyamoto2015pluripotency, huang2020decoding}.
Here, $\tau$ denotes the characteristic timescale for epigenetic modifications, which is assumed to be sufficiently larger than 1; the change in epigenetic modification is much slower than that of gene regulatory dynamics \cite{barth2010fast, maeshima2015physical, sasai2013time}.

Recalling the relevance of oscillatory dynamics, we chose a GRN in which oscillatory dynamics were generated for appropriate $\theta_i$ values (specifically at $\theta_i \sim 0$).
First, we adopted a repressilator model as a minimal model (see Fig. S4a \footnote{See Supplemental Information for derivations, detailed discussions, and figures.}),
consisting of three genes that repress the expression of the next gene in a cyclic manner \cite{elowitz2000synthetic}.
Specifically, we chose $J_{21} = J_{32} = J_{13} =-g = - 0.4$ in Eq. (\ref{dx}).

\begin{figure}[tbp]
    \centering
        \includegraphics[width=\linewidth]{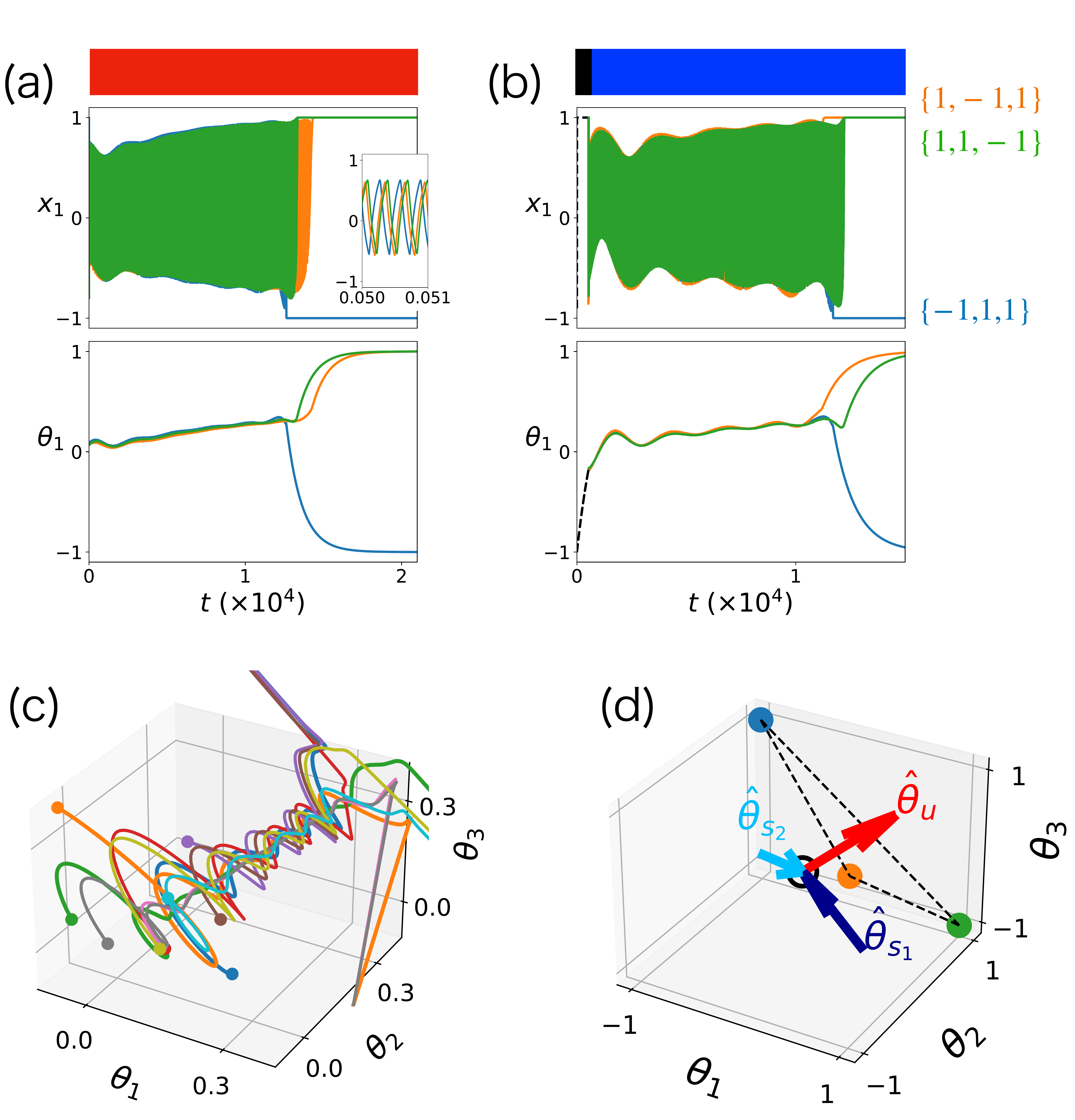}
    % {\large (a) \hspace{0.2\linewidth} (b)\vspace{-0.3cm}}
    % \subfigure%[]
    % {
    %     \includegraphics[width=0.9\linewidth]{Figure/ex_rep.pdf}
    % }
    % \vspace{-0.5cm}
    % \setcounter{subfigure}{2}
    % \subfigure[]
    % {
    %     \includegraphics[width=0.46\linewidth]{Figure/3t.pdf}
    % }
    % \subfigure[]
    % {
    %     \includegraphics[width=0.46\linewidth]{Figure/eigen_vector.pdf}
    % }
    \caption
    {
    (a) Cell differentiation and (b) reprogramming of the repressilator model with $\tau = 10^3$.
    Upper bar indicates differentiation (red), reprogramming manipulation (black), and later process (blue) without it ($I_i = 0$).
    We plotted the time development of $x_1, \theta_1$ (see Fig. S5ab\cite{Note2} for the time series of all variables).
    (a) Three trajectories are sampled from slightly different initial conditions near $\theta_i = 0$.
    (b) From the fixed point $\{-1, 1, 1\}$, we tested three slightly different time spans to add external input (520, 530, 540).
    After reprogramming manipulation, the cellular state first approached $\theta_1 = \theta_2 = \theta_3$, and then differentiated to three fixed points again.
    (c) Trajectories through reprogramming and differentiation plotted in the $(\theta_1, \theta_2, \theta_3)$ space.
    Ten attempts were overlaid by considering the initial conditions in $x_i \in [-1, 1], \theta_i \in [-0.25, 0.25]$, which allowed oscillation.
    Attraction for the straight line $\theta_1 = \theta_2 = \theta_3$ and departure from it was discernible. 
    (d) Eigenvector $\{\boldsymbol{v}_k\}$ and variables $\{\hat{\theta}_{k}\}$ (see text).
    Three colored points represent the fixed points $\{-1, 1, 1\}$ (blue), $\{1, -1, 1\}$ (orange), and $\{1, 1, -1\}$ (green).
    }
    \label{x-theta}
    \vspace{-0.4cm}
\end{figure}

The expression of $x_i$ in this model showed a limit-cycle oscillation when $\theta_i$ was close to zero.
Thus, for the epigenetic modification to change $\theta_i$ following Eq. (\ref{dtheta}), the states were differentiated into three fixed-point attractors $\{\theta_1, \theta_2, \theta_3\} = \{-1, 1, 1\}, \{1, -1, 1\}, \{1, 1, -1\}$, after first approaching a straight line $\theta_1 = \theta_2 = \theta_3$, as shown in Fig. \ref{x-theta}a 
\footnote{Considering positive/negative symmetry, six attractors exist in the whole space.
In this letter, we considered only the side $\sum \theta_i > 0$.} (see also Fig. S5a\cite{Note2}).
In these fixed points, $d \theta_i / dt = 0$ were satisfied; that is, the differentiation of expression $x_i$ was embedded into the epigenetic modification $\theta_i$.

Now, we considered "reprogramming."
Starting from one of the differentiated fixed points, we added external input $I_i(t)$ to invoke the transient oscillation again (black dotted line in Fig. \ref{x-theta}b).
Later, $I_i(t)$ was set at zero.
After reprogramming manipulation, they approached a line with $\theta_1 = \theta_2 = \theta_3$ around the origin, and then deviated from the line to one of the three fixed points (Fig. \ref{x-theta}b), in the same manner as the differentiation process.
During this reprogramming process, memory of the differentiated states was erased.
Once the oscillation in $x$ was recovered, the approach to the straight line and deviation from it always followed (Fig. \ref{x-theta}c).

Next, we studied how the attraction to the straight line first occurs, followed by differentiation progression.
For it, we considered the adiabatic limit of $\tau \rightarrow \infty$.
For fixed $\theta_i$, we first obtained the attractor $x_i$.
Then, the evolution of $\theta_i$ was obtained by replacing $x_i$ in Eq. (\ref{dtheta}) by its time average $\bar{x}_i$ for a given $\theta_i$, as follows:
\begin{align}
    \frac{d \theta_i}{d t} = \bar{x}_i(\theta_i) - \theta_i (\equiv \Theta_i). \label{ae_dtheta}
\end{align}
In the three-variable Eq. (\ref{ae_dtheta}), $\{\theta_1, \theta_2, \theta_3\} = \{0, 0, 0\}$ is a fixed point solution because $x_i(t)$ showed a symmetric limit-cycle oscillation, such that $\bar{x}_i=0$ for all $i$ therein for $\{\theta_1, \theta_2, \theta_3\}=\{0, 0, 0\}$.
By slightly perturbing $\theta_i$ as a parameter, $\bar{x}_j$ changed accordingly.
From $\partial \bar{x}_i / \partial \theta_j$, we obtained the Jacobi matrix $\partial \Theta_i / \partial \theta_j$ with eigenvalues $\{ \lambda_k \}$ and eigenvectors $\{ \boldsymbol{v}_k \}$.
As shown in Fig. \ref{x-theta}d, $\{\theta_1, \theta_2, \theta_3\} = \{0, 0, 0\}$-fixed point was a saddle, with the eigenvector $\boldsymbol{v}_{u} = \{1, 1, 1\}/\sqrt{3}$ corresponding to $\lambda_{u} > 0$ (unstable axis), and $\boldsymbol{v}_{s_1} = \{2, -1, -1\}/\sqrt{6}, \boldsymbol{v}_{s_2} = \{0, -1, 1\}/\sqrt{2}$ for $\mathrm{Re}(\lambda_{s_1}) = \mathrm{Re}(\lambda_{s_2}) < 0$ (see Supplemental Information (SI) 1A\cite{Note2}).
To investigate $\theta$ dynamics along each eigenvectors $\boldsymbol{v}_{k}$ ($k = u, s_1, s_2$), we then introduced the variable $\hat{\theta}_k$, projection of $\boldsymbol{\theta}$ on $\boldsymbol{v}_k$ (that is, $\hat{\theta}_k = \boldsymbol{\theta} \cdot \boldsymbol{v}_k$, with $\boldsymbol{v}_k$ normalized).
Noteworthy, owing to the symmetry of the repressilator, the unstable manifold was in line with the eigenvector $\boldsymbol{v}_{u}$ (see SI 1A\cite{Note2}).

\begin{figure}[tbp]
    \centering
        \includegraphics[width=\linewidth]{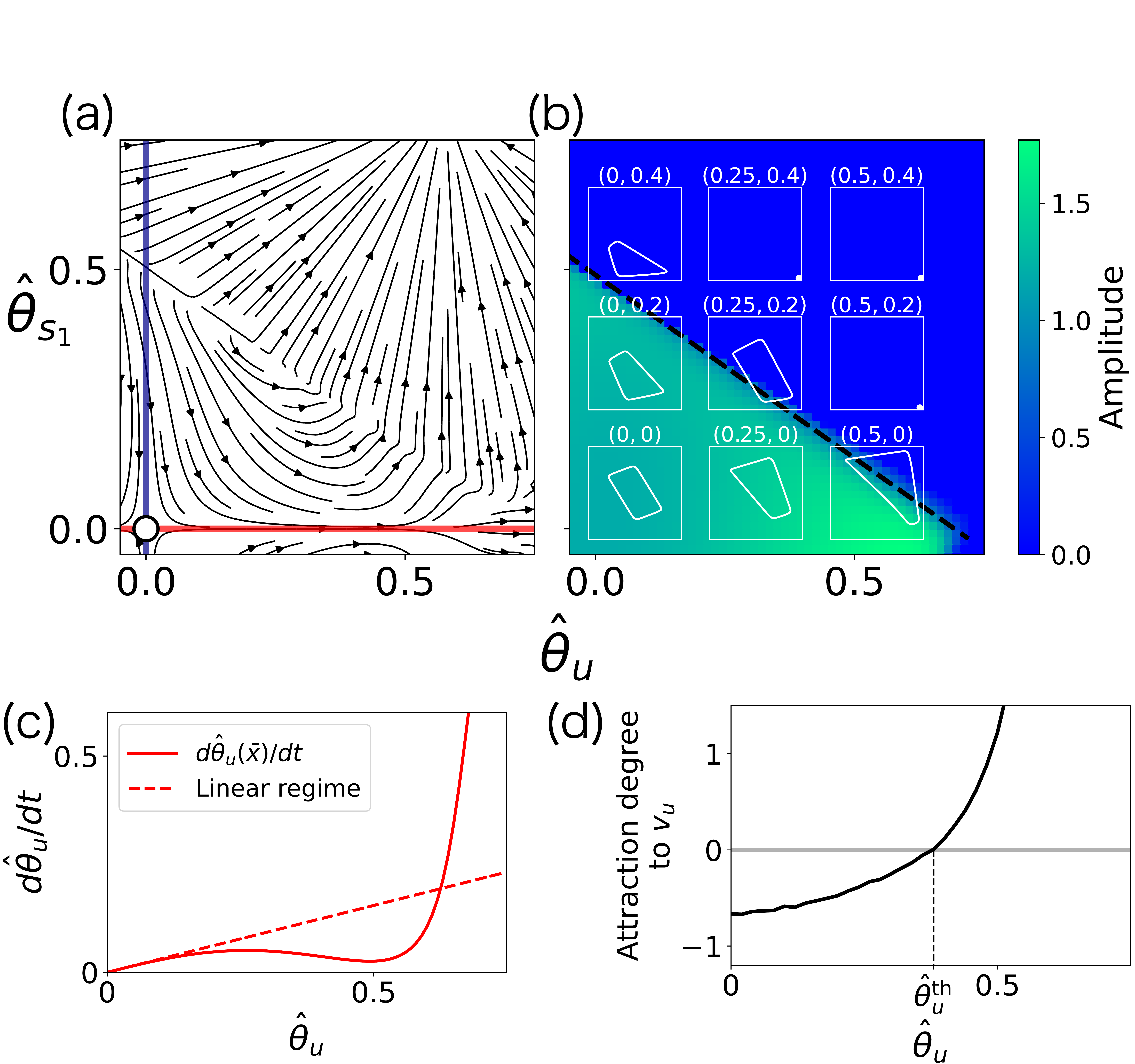}
    % {\large (a) \hspace{0.3\linewidth} (b)\vspace{-0.3cm}}
    % \vspace{-0.3cm}
    % \subfigure%[]
    % {
    %     \includegraphics[width=0.96\linewidth]{Figure/pd_flow.pdf}
    % }
    % \vspace{-0.5cm}
    % \setcounter{subfigure}{2}
    % \subfigure[]
    % {
    %     \includegraphics[width=0.46\linewidth]{Figure/v3.pdf}
    % }
    % \subfigure[]
    % {
    %     \includegraphics[width=0.46\linewidth]{Figure/stability.pdf}
    % }
    \caption
    {
    (a) Stream plot of $(d \hat{\theta}_{u}/dt, d \hat{\theta}_{s_1}/dt)$ in $(\hat{\theta}_{u}, \hat{\theta}_{s_1})$ space according to Eq. (\ref{ae_dtheta}).
    Red (blue) line represents the direction of the eigenvector $\boldsymbol{v}_{u} (\boldsymbol{v}_{s_1})$.
    (b) Attractors in the $x$-space for each fixed $\theta$ value.
    For the green (blue)-colored region, the attractor in the $x$ space was the limit cycle (fixed point).
    Each trajectory shows the dynamics in the $x_1-x_2$ space for $(\hat{\theta}_{u}, \hat{\theta}_{s_1})$ (both $x_1$ and $x_2$ axes for all figures are ranged within $\{-1.05, 1.05\}$).
    (c) $d\hat{\theta}_{u} / d t$ plotted as a function of $\hat{\theta}_{u}$.
    For comparison, we plotted $d \hat{\theta}_{u} = \lambda_{u} \hat{\theta}_{u}$(red dotted line).
    (d) Degree of attraction to $\boldsymbol{v}_{u}$.
    $\nu$, in the text, is plotted as a function of $\hat{\theta}_{u}$.
    If it is negative (positive), $\{\theta\}$ was attracted to (departed from) $\boldsymbol{v}_{u}$.
    See Fig. S6\cite{Note2} for more detail.
    }
    \label{analysis-rep}
    \vspace{-0.5cm}
\end{figure}

As shown in Fig. \ref{x-theta}cd, the straight line to which all trajectories converged agreed with the unstable manifold $\boldsymbol{v}_{u}$ (Fig. \ref{analysis-rep}a).
Of course, attraction to the $\boldsymbol{v}_{u}$ axis was natural if the initial conditions were restricted to the stable manifold for $\{\theta_1, \theta_2, \theta_3\}=\{0, 0, 0\}$.
However, we observed an attraction toward the unstable axis over a wide range of initial conditions for $\{\theta_i\}$, which supports the oscillation of $x_i$. 
Furthermore, the magnitudes of the eigenvalues for the stable and unstable eigenvectors were of the same order ($\mathrm{Re}(\lambda_{u}) = 0.31, \mathrm{Re}(\lambda_{s_1}) = \mathrm{Re}(\lambda_{s_2}) = -0.66$, see Fig. S5c\cite{Note2}).
Thus, the reprogramming dynamics shown in Fig. \ref{x-theta}b could not be explained just by the linear stability.

To elucidate whether the nonlinear effect suppresses the instability along the $\boldsymbol{v}_{u}$ axis, we computed $d \hat{\theta}_{u} / d t$.
As shown in Fig. \ref{analysis-rep}c, $d \hat{\theta}_{u}/ d t$ was drastically reduced from the linear case.
We also computed $(d \hat{\theta}_{s_1} / d t, d \hat{\theta}_{s_2} / d t)$ for a certain $\hat{\theta}_{u}$ value (that is, the flow structure in the $(\hat{\theta}_{s_1}, \hat{\theta}_{s_2})$ plane, sliced along the $\hat{\theta}_{u}$ axis), which showed that $\hat{\theta}_{s_1} = 0$ changed from stable to unstable at $\hat{\theta}_{u} = \hat{\theta}_{u}^{\mathrm{th}} (\sim0.4)$ (see Fig. \ref{analysis-rep}d and Fig. S6\cite{Note2}).
Up to $\hat{\theta}_{u} < \hat{\theta}_{u}^{\mathrm{th}} (\sim 0.4)$, $\theta_i$ in $(\hat{\theta}_{s_1}, \hat{\theta}_{s_2})$ plane was attracted to the $\hat{\theta}_{u}$ axis.
By further increasing $\hat{\theta}_{u}$ beyond $\hat{\theta}_{u}^{\mathrm{th}}$, $\theta_i$ departured from the $\hat{\theta}_{u}$ axis rotating in the $(\hat{\theta}_{s_1}, \hat{\theta}_{s_2})$ plane, leading to differentiation into three distinct fixed points.

To unveil how the slow motion along $\hat{\theta}_{u}$ and the attraction to $\hat{\theta}_{u}$ occurred, we first fixed $\theta_i$ and studied the change in the $x$ attractor, as shown in Fig. \ref{analysis-rep}b.
In the green (blue) region, the $x$ attractor was a limit cycle (fixed point) for $(\hat{\theta}_{u}, \hat{\theta}_{s_1})$.
At the line $\hat{\theta}_{s_1} = -\hat{\theta}_{u}/\sqrt{2} + \sqrt{6}/5$ (as discussed in SI 1BC\cite{Note2}), $x$-dynamics exhibited bifurcation from the limit cycle to a fixed point $\{1, -1, 1\}$ (see Fig. S7\cite{Note2} for more details).
Considering the symmetry of the repressilator, bifurcations to three fixed points $\{-1, 1, 1\}, \{1, -1, 1\}, \{1, 1, -1\}$ coexisted in the $(\hat{\theta}_{s_1},\hat{\theta}_{s_2})$ plane.
With the increase of $\hat{\theta}_{u}$, the limit cycle approached the three fixed points.

\begin{figure}[tbp]
    \centering
        \includegraphics[width=\linewidth]{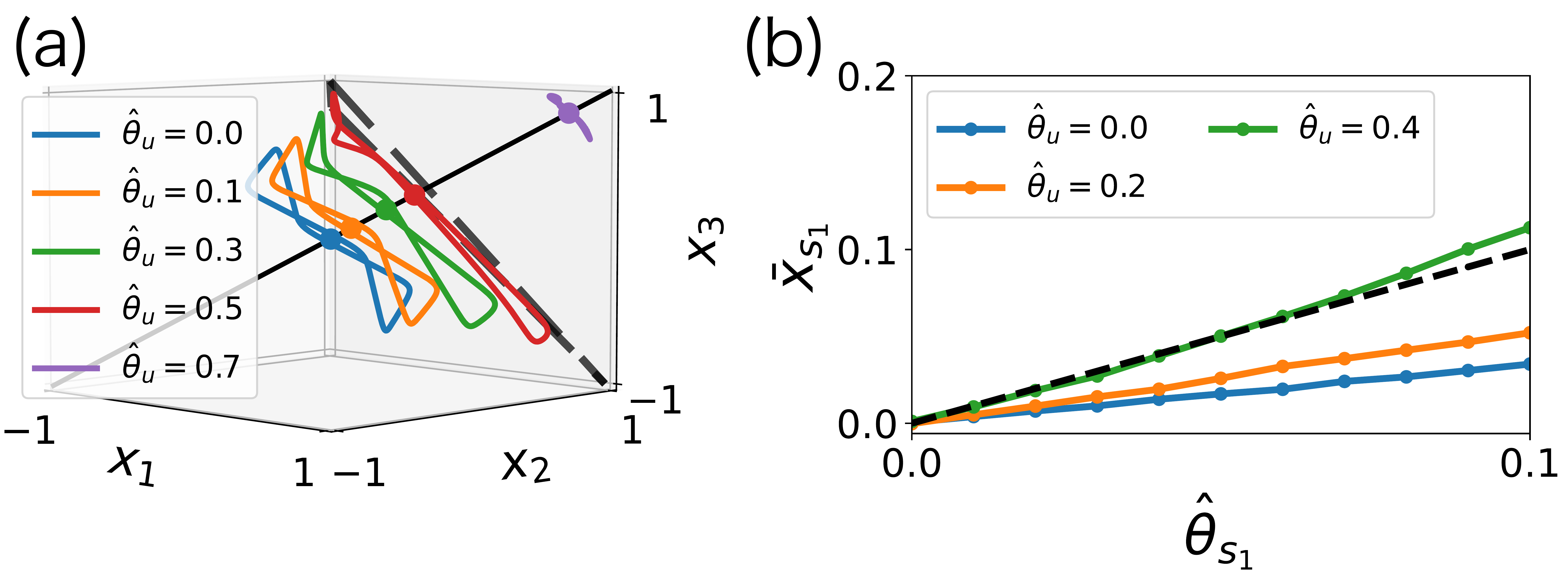}
    % \subfigure[]
    % {
    %     \includegraphics[width=0.45\linewidth]{Figure/3x_v3.pdf}
    % }
    % \vspace{-0.3cm}
    % \subfigure[]
    % {
    %     \includegraphics[width=0.45\linewidth]{Figure/bx_s1.pdf}
    % }
    \caption
    {
    (a) Change in the limit cycle $x(t)$ (closed line) and $\bar{x}$ (point) along $\hat{\theta}_{u}$.
    Black dotted line shows an equilateral triangle with corners $\{-1, 1, 1\}, \{1, -1, 1\}, \{1, 1, -1\}$ (see SI 1D\cite{Note2}).
    (b) $\bar{x}_{s_1}$ as a function of $\hat{\theta}_{s_1}$.
    The slope of each line corresponds to $\partial \bar{x}_{s_1}/\partial \hat{\theta}_{s_1}$ for $\hat{\theta}_{u}$.
    From $\partial \Theta_{s_1}/d \hat{\theta}_{s_1}$, if the slope is less than one, the orbits are attracted toward $\hat{\theta}_{u}$.
    For comparison, we plotted $\bar{x}_{s_1} = \hat{\theta}_{s_1}$ as a black dotted line.
    At $\hat{\theta}_{u}\sim 0.4$, $\partial \bar{x}_{s_1}/\partial \hat{\theta}_{s_1}$ exceeds one.
    }
    \label{x_v3}
    \vspace{-0.5cm}
\end{figure}

Now, we discussed the mechanism of slow motion along $\hat{\theta}_{u}$.
From Eq. (\ref{ae_dtheta}), movement along $\boldsymbol{v}_{u}$ followed $d \hat{\theta}_{u} /d t = \bar{x}_{u}(\hat{\theta}_{u}) - \hat{\theta}_{u}$ (we defined $x_k$ as a projection on $\boldsymbol{v}_k$).
As shown in Fig. \ref{x_v3}a, the limit cycle approached the plane spanned by the three fixed points $\{-1, 1, 1\}, \{1, -1, 1\}, \{1, 1, -1\}$ as $\hat{\theta}_{u}$ increased.
In the plane, $\bar{x}_u$ comprised $1/\sqrt{3}$, and $\theta_i$ increased following Eq. (\ref{ae_dtheta}) (see SI 1D\cite{Note2} for more details).
Then, as $\hat{\theta}_{u}$ approaches $1/\sqrt{3}$, $d \hat{\theta}_{u}/d t$ was minimized, as shown in Fig. \ref{analysis-rep}c.

Next, we considered how attraction to the $\boldsymbol{v}_{u}$ from the $(\hat{\theta}_{s_1}, \hat{\theta}_{s_2})$ plane was lost at the $\hat{\theta}_{u} =\hat{\theta}_{u}^{th}$.
By considering $\hat{\theta}_{u}$ as a parameter, the direction of flow in the $(\hat{\theta}_{s_1}, \hat{\theta}_{s_2})$ plane toward the $\boldsymbol{v}_{u}$ was determined by the sign of $\partial \Theta_{s_1} / \partial \hat{\theta}_{s_1} = \partial \bar{x}_{s_1}/\partial \hat{\theta}_{s_1} - 1 \equiv \nu \Theta_{s_1}$ (we defined $\Theta_{s_1}$ as a projection on $\boldsymbol{v}_{s_1}$).
As shown in Fig \ref{analysis-rep}b, with the increase in $\hat{\theta}_{u}$, the bifurcation point from the limit cycle to the fixed point approached the $\hat{\theta}_{u} = 0$ line.
Hence, by slightly changing $\hat{\theta}_{s_1}$, $\bar{x}$ reached fixed points.
Accordingly, $\partial \bar{x}_{s_1}/\partial \hat{\theta}_{s_1}$ increase beyond one, so that $\partial \Theta_{s_1}/\partial \hat{\theta}_{s_1}$ became positive at $\hat{\theta}_{u}$, approaching $\hat{\theta}_{u}^{\mathrm{th}} \sim 0.4$, as shown in Figs. \ref{analysis-rep}d and \ref{x_v3}b.

Thus, we have unveiled how attraction to the unstable manifold is achieved by slow epigenetic fixation of the oscillation of fast gene expression in the repressilator model.
Following this picture, reprogramming is possible by forcing the cells to return to the oscillatory state.
Then, the cell is attracted to a pluripotent state with low epigenetic modification $\theta_i \sim 0$, from which differentiation to distinct cell types with specific $\theta$ values follows.

To verify the generality of this reprogramming scheme, we examined several GRN models with more degrees of freedom.
As discussed in \cite{matsushita2020homeorhesis}, differentiation from oscillatory states is often observed in GRNs (e.g., 20\% of randomly generated GRNs show oscillatory dynamics for $N = 10$).
An example is shown in Fig. S8a\cite{Note2}.
From a differentiated state, we overexpressed three genes to regain oscillatory expression (black line in Fig. S8a\cite{Note2}).
Later, global attraction to unstable manifold also occurred as discussed above.
Then, the cell states branch to distinct fixed point states again (blue line in Fig. S8a\cite{Note2}).
In these cases, the original pluripotent state with $\theta=0$ was an unstable fixed point, with one positive eigenvalue for the Jacobi matrix of $\theta$ dynamics (Fig. S8d\cite{Note2}), as in the repressilator model.
Even though the degrees of freedom increased, the unstable manifold is one-dimensional, and the attraction to the manifold occurred from a higher-dimensional state space.
This implies that reprogramming manipulation requires only partial degrees of freedom compared with the total number of genes.
In fact, overexpression of three genes is sufficient for reprogramming in GRN models with $N = 10$, as far as we have investigated.

\begin{figure}[tbp]
\centering
        \includegraphics[width=\linewidth]{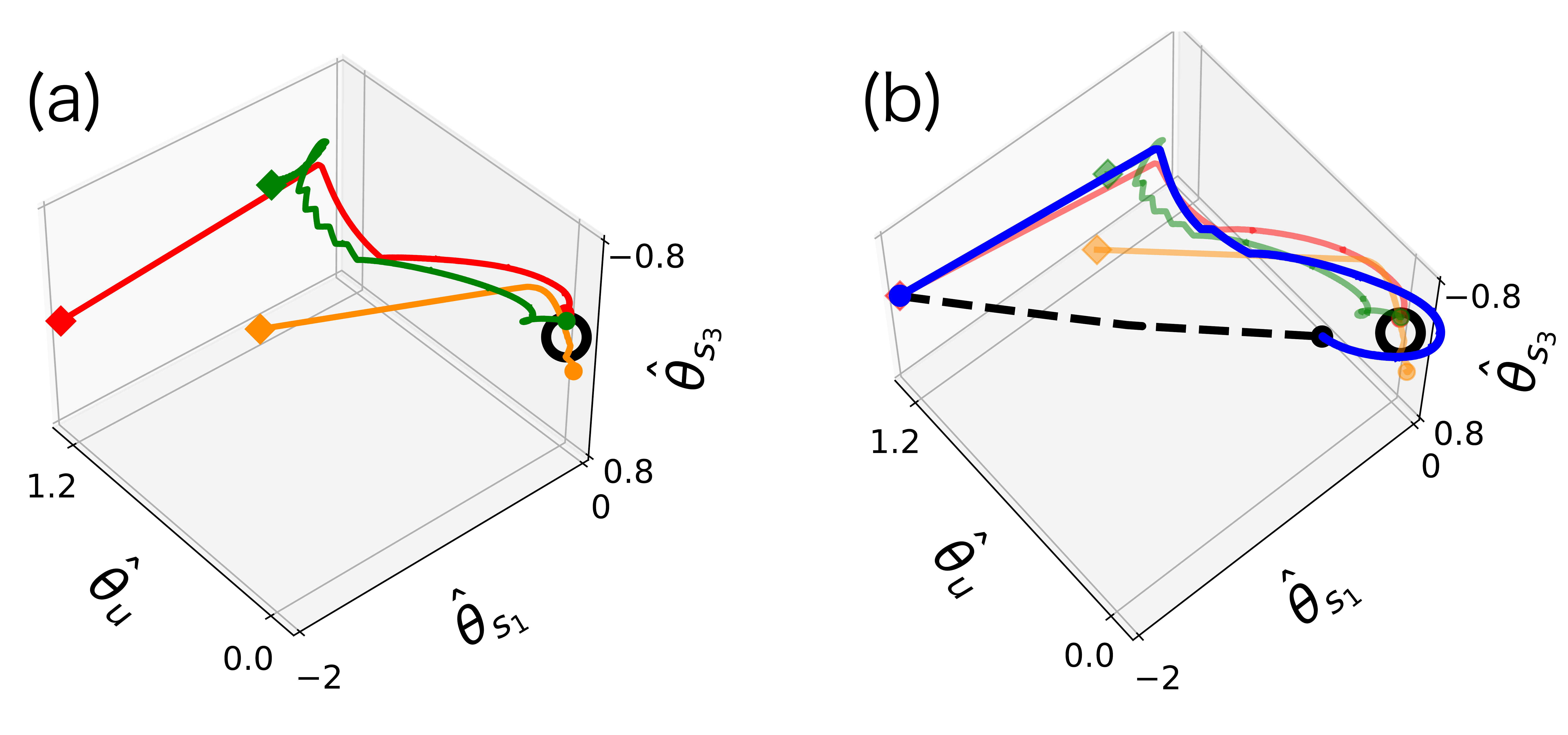}
        % \subfigure[]
        % {
        %     \includegraphics[width=0.45\linewidth]{Figure/3t5g.pdf}
        % }
        % \vspace{-0.3cm}
        % \subfigure[]
        % {
        %     \includegraphics[width=0.45\linewidth]{Figure/3t5gr.pdf}
        % }
    \caption
    {
    (a) Cell differentiation and (b) reprogramming in the five-gene model
    (a) Three orbits starting from the vicinity of the saddle point $\theta_i = 0$ for all $i$ (black dotted point), reached three distinct cell types.
    (b) From differentiated cell types (red point), we added external input $I_i(t)$ to Nanog, Oct4, and Klf4 for a certain time span (black dotted line).
    After such reprogramming manipulation, we set $I_i(t) = 0$.
    The cell state then spontaneously approached the saddle point, and the reinitiated the differentiation progression again (blue line).
    $\tau = 10^{3}$.
    See Fig. S9\cite{Note2} for more detail.
    }
    \label{5g}
    \vspace{-0.3cm}
\end{figure}

The present mechanism also works in a model extracted from GRN for an ES cell \cite{dunn2014defining}, as a core network with five genes (\textit{Nanog, Oct4, Gata6, Gata4,} and \textit{Klf4}) \cite{miyamoto2015pluripotency} (see Fig. S4b).
\textit{Oct4, Sox2}, and \textit{Klf4} are known as factors to induce reprogramming.
The model involves a negative feedback loop, as in the repressilator, in addition to positive feedback regulation.
In this five-gene model, $x_i$ and $\theta_i$ oscillate in the region near the origin, and then differentiation to three fixed points progresses as in the repressilator case (three lines in Fig. \ref{5g}a), whereas $\theta_i = 0$ for all $i$ represents a saddle point with one unstable manifold and four stable manifolds, as shown in Fig.S9b.
After overexpression of \textit{Oct4}, \textit{Nanog}, and \textit{Klf4} in one of the differentiated cell types for a certain time span (black dotted line in Fig. \ref{5g}b)\footnote{\textit{Sox2} is reduced into Nanog in the five-gene model}, the epigenetic state $\theta_i$ approached the unstable manifold for the unstable fixed point $\theta_i = 0$, leading to recovery of pluripotency (blue line in Fig. \ref{5g}b).

In this letter, we have shown that oscillatory gene expression dynamics with slow epigenetic modifications lead to cellular reprogramming by overexpression of only few genes.
The global attraction to the unstable manifold of the saddle point explains the reprogramming process.
Now, the return to the top of the landscape by reprogramming, which is seemingly unstable, is explained by the strong attraction toward the unstable manifold of the saddle, and suppressed instability along with the unstable manifold, owing to the approach of limit-cycle of bifurcation to fixed points.
The memory of the cellular state before reprogramming manipulation was erased through this reprogramming process.

Moreover, regain of oscillation was found to be the main requirement for reprogramming, whereas elaborate manipulations to induce a cellular state into specific states is not necessary.
This explains the role of oscillations in gene expression in pluripotent cells \cite{kobayashi2009cyclic} and epigenetic modification through the differentiation process \cite{rulands2018genome}, as well as it explains how reprogramming is possible by overexpressing just few genes among thousand of that \cite{takahashi2006induction, velychko2019excluding}.
The timescale separation between fast expression dynamics and slow epigenetic modification feedback required is also consistent with previous observations \cite{barth2010fast, maeshima2015physical}.
In future studies, experimental support is necessary, as well as theoretical analysis of slow-fast dynamical systems \cite{aoki2013slow, kuehn2015multiple}.

\begin{acknowledgments}
This research was partially supported by a Grant-in-Aid for Scientific Research on Innovative Areas (17H06386) from the Ministry of Education, Culture, Sports, Science, and Technology of Japan, and a Grant-in-Aid for Scientific Research (A) (20H00123) from the Japanese Society for the Promotion of Science.
\end{acknowledgments}
\bibliographystyle{apsrev4-2}
% \bibliography{apssamp}% Produces the bibliography via BibTeX.
%

\end{document}

\preprint{AIP/123-QED}

\title[Supplemental Information]{Supplemental Information:\\ Dynamical-systems theory of cellular reprogramming}% Force line breaks with \\
% \thanks{Footnote to title of article.}

\author{Yuuki Matsushita}
\author{Tetsuhiro S. Hatakeyama}%
\author{Kunihiko Kaneko}
% \email{kaneko@complex.c.u-tokyo.ac.jp}
% \affiliation{%
% Department of Basic Science, Graduate School of Arts and Sciences, University of Tokyo, 3-8-1 Komaba, Meguro-ku, Tokyo 153-8902, Japan}

\date{\today}% It is always \today, today,

\maketitle

\renewcommand{\thefigure}{S\arabic{figure}}
\setcounter{figure}{0}
\renewcommand{\theequation}{S\arabic{equation}}
\setcounter{equation}{0}

\section{Analysis of the repressilator}
\subsection{Derivation of eigenvalues and eigenvectors for the adiabatically-eliminated $\theta$ dynamics}
We considered the Jacobi matrix of Eq.(2) in the repressilator model as
\begin{align}
    \frac{\partial \Theta_i}{\partial \theta_j} = \frac{\partial \bar{x}_i}{\partial \theta_j} - \delta_{ij}
    (i, j = 1, 2, 3), \label{tJacobiMatrix}
\end{align}
and fixed-point analysis around $\{\theta_1, \theta_2, \theta_3\} = \{0, 0, 0\}$.
The repressilator model obviously has symmetry against a transformation in $1 \rightarrow 2 \rightarrow 3 \rightarrow 1$ around $\{\theta_1, \theta_2, \theta_3\} = \{0, 0, 0\}$.
\begin{align}
&\frac{\partial \bar{x}_1}{\partial \theta_1} = \frac{\partial \bar{x}_2}{\partial \theta_2} = \frac{\partial \bar{x}_3}{\partial \theta_3} (=a_0), \\
&\frac{\partial \bar{x}_2}{\partial \theta_1} = \frac{\partial \bar{x}_3}{\partial \theta_2} = \frac{\partial \bar{x}_1}{\partial \theta_3} (=a_{+1}), \\
&\frac{\partial \bar{x}_3}{\partial \theta_1} = \frac{\partial \bar{x}_1}{\partial \theta_2} = \frac{\partial \bar{x}_2}{\partial \theta_3} (=a_{-1}).
\end{align}
Eq. (\ref{tJacobiMatrix}) can be written as the following circulant matrix:
\begin{align}
\frac{\partial \Theta}{\partial \theta} =
\begin{bmatrix}
a_0 - 1 & a_{-1} & a_{+1} \\
a_{+1} & a_0 - 1 & a_{-1} \\
a_{-1} & a_{+1} & a_0 - 1
\end{bmatrix}.
\end{align}

Here, we considered the following matrix $C$:
\begin{align}
C = 
\begin{bmatrix}
0 & 0 & 1 \\
1 & 0 & 0 \\
0 & 1 & 0
\end{bmatrix}.
\end{align}

$\partial \Theta/\partial \theta$ can be expanded as:
\begin{align}
    \frac{\partial \Theta}{\partial \theta} = (a_0 - 1) C^0 + a_{+1} C + a_{-1} C^2.
\end{align}

Matrix $C$ can be diagonalized by a following $3\times3$ discrete Fourier transform (DFT) matrix $F$:

\begin{align}
F = 
\begin{bmatrix}
(\omega_0)^0 & (\omega_1)^0 & (\omega_2)^0 \\
(\omega_0)^1 & (\omega_1)^1 & (\omega_2)^1 \\
(\omega_0)^2 & (\omega_1)^2 & (\omega_2)^2
\end{bmatrix},
\end{align}
where $\omega_j = \exp(2\pi ij/3)$ with $j = 0, 1, 2$ ($i$, used here, is an imaginary unit).
Thus, the circulant matrix $\partial \Theta/\partial \theta$ can be diagonalized using the DFT matrix $F$.

The $j$th eigenvector $\boldsymbol{v}_j$ is now written as:
\begin{align}
\boldsymbol{v}_j = \{(\omega_j)^0, (\omega_j)^1, (\omega_j)^2\}.
\end{align}

Each eigenvalue $\lambda_j$ corresponding to the eigenvector $\boldsymbol{v}_j$ is
\begin{align}
\lambda_j = (a_0 - 1) + a_{-1}\omega_j^1 + a_{+1} \omega_j^2.
\end{align}

Accordingly, we obtained
\begin{align}
\lambda_0 &= (a_0 - 1) + a_{+1} + a_{-1}, \\
\lambda_1, \lambda_2 &= (a_0 - 1) -\frac{a_{+1} + a_{-1}}{2} \pm \frac{\sqrt{3}i}{2}(a_{-1} - a_{+1}).
\end{align}

Now, we estimated the value of $a_0, a_{+1}, a_{-1}$.
$a_0 = \partial \bar{x}_i/\partial \theta_i$ is given by the change in gene expression against the change in epigenetic modification of the same gene (see Fig. \ref{Model}a).
In our model, we adopted a positive feedback relationship between gene expression and epigenetic modification of the gene $i$; $a_0$ is positive.
In contrast, $a_{+1}$ is given by the change in gene expression against the change in epigenetic modification of the repressing gene; $a_{+1}$ is negative.
Similarly, $a_{-1}$ is positive.
In addition, the change in gene expression was amplified throughout the negative repressilator loop; thus, $a_0, a_{+1}, a_{-1}$ should satisfy
\begin{align}
    |a_0| < |a_{+1}| < |a_{-1}|.
\end{align}
Indeed, from numerical calculations, the values of $a_0, a_{+1}, a_{-1}$ followed these estimations ($a_0 = 0.66, a_{+1} = -1.08, a_{-1} = 1.72$ for $g = -0.4$).
From these estimations, we obtained the relation of eigenvalues as:
\begin{align}
\mathrm{Re}(\lambda_1) = \mathrm{Re}(\lambda_2) < \mathrm{Re}(\lambda_0)
\end{align}

In the present case ($a_0 = 0.66, a_{+1} = -1.08, a_{-1} = 1.72$ for $g = -0.4$), $\lambda_0$ is positive, and $\lambda_1, \lambda_2$ is negative:
\begin{align}
\mathrm{Re}(\lambda_1) = \mathrm{Re}(\lambda_2) < 0 < \mathrm{Re}(\lambda_0)
\end{align}
Thus, $\boldsymbol{\theta} = \boldsymbol{0}$-saddle point consists of an eigenvector $\boldsymbol{v}_{0}$ with $\lambda_0 > 0$ and eigenvectors $\boldsymbol{v}_{0}, \boldsymbol{v}_1$ with $\mathrm{Re}(\lambda_1) = \mathrm{Re}(\lambda_2) < 0$.
To analyze and plot in real $\theta$ space, we adopted $\boldsymbol{v}_{u} \propto \boldsymbol{v}_0, \boldsymbol{v}_{s_1} \propto (\boldsymbol{v}_1+\boldsymbol{v}_2)/2, \boldsymbol{v}_{s_2} \propto (\boldsymbol{v}_1-\boldsymbol{v}_2)/2i$ with a normalization factor.
The symmetry of the repressilator against the transformation in $1 \rightarrow 2 \rightarrow 3 \rightarrow 1$ is not broken as long as $\theta_1 = \theta_2 = \theta_3$ is satisfied.
Therefore, the direction of the eigenvector $\boldsymbol{v}_{u}$ always coincides with that of the unstable manifold.

\subsection{Linear stability analysis for $x$ dynamics with fixed $\theta$}\label{FixedPointAnalysis}
We discussed $x$ dynamics of repressilator model with fixed $\theta$ as parameter.
Full model with fixed $\theta$ can be written as
\begin{subequations}
\begin{align}
    \frac{d x_1}{d t} &= \tanh \{ -\beta g(x_3 - \theta_1/g)\} - x_1 (= X_1) \label{X1}\\
    \frac{d x_2}{d t} &= \tanh \{ -\beta g(x_1 - \theta_2/g)\} - x_2 (= X_2) \label{X2}\\
    \frac{d x_3}{d t} &= \tanh \{ -\beta g(x_2 - \theta_3/g)\} - x_3 (= X_3). \label{X3}
\end{align}
\end{subequations}
Thus, Jacobi Matrix $\partial X_i/\partial x_j$ can also be written as
\begin{widetext}
        \begin{align}
\frac{\partial X}{\partial x} =
\begin{bmatrix}
-1 & 0 & \frac{-\beta g}{\cosh^2(-\beta g (x_3 - \theta_1/g))} \\
\frac{-\beta g}{\cosh^2(-\beta g (x_1 - \theta_2/g))}& -1 & 0 \\
0 & \frac{-\beta g}{\cosh^2(-\beta g (x_2 - \theta_3/g))}& -1
\end{bmatrix}. \label{JacobiMatrix}
         \end{align}
\end{widetext}
First, we studied the stability of the fixed point for $\{\theta_1, \theta_2, \theta_3\} = \{0, 0, 0\}$ to elucidate the condition for oscillation in $x$ dynamics with no epigenetic modification.
In this case, $\{x_1, x_2, x_3\} = \{0, 0, 0\}$ satisfies the fixed-point condition.

By inserting $x_i = 0, \theta_i = 0$ for all $i$ in Eq. (\ref{JacobiMatrix}), we obtained:
\begin{align}
\frac{\partial X}{\partial x} = \begin{bmatrix}
-1 & 0 & -\beta g \\
-\beta g & -1 & 0 \\
0 & -\beta g & -1
\end{bmatrix}. \label{JacobiMatrix0}
\end{align}

From Eq. (\ref{JacobiMatrix0}), eigenvalue $\lambda$ is given by
\begin{align}
    \lambda = -\beta g - 1, (\frac{1}{2}\beta g - 1) \pm \frac{\sqrt{3}}{2}\beta g i. \label{eigen_value_x}
\end{align}
In the repressilator model($g > 0$), the first value of (\ref{eigen_value_x}) is always negative.
The fixed point is unstable for 
\begin{align}
    g > \frac{2}{\beta},
\end{align}
where the limit-cycle attractor appeared as a result of Hopf bifurcation, 
to specific $g > 0.05$ for $\beta = 40$.

\begin{figure}[tbp]
    \centering
    \subfigure%[]
    {
        \includegraphics[width=0.5\linewidth]{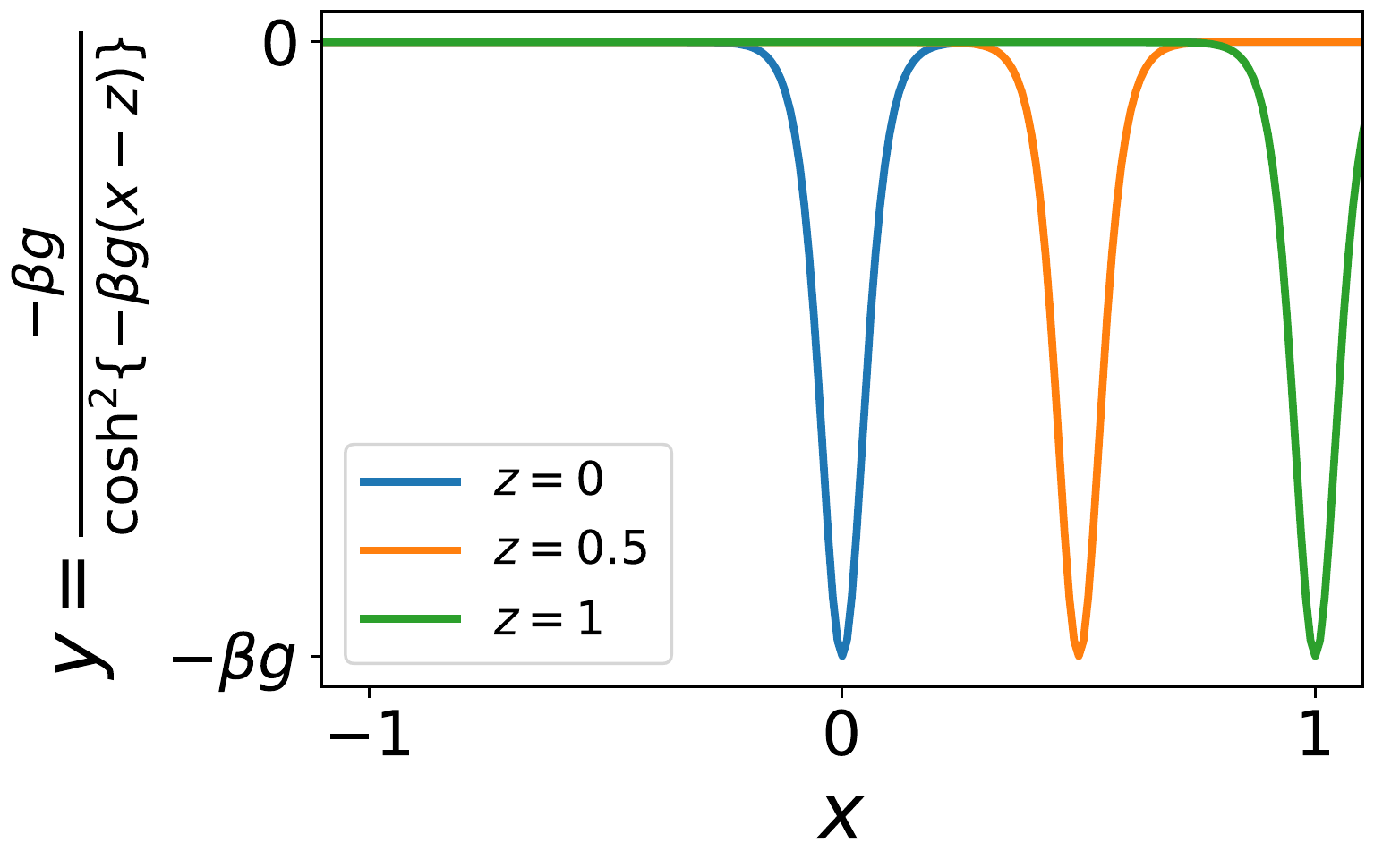}
    }
    \caption
    {
    $-\beta g/\cosh^2(\beta g(x - z))$ as a function of $x$ ($\beta = 40, g = 0.4$).
    This function has a tall spike at $x = z$.
    Considering the range of the $\tanh$ function, this term could be regarded as zero if $|z| > 1$ for a sufficiently large $\beta$.
    }
    \label{cosh}
\end{figure}

Next, we considered the bifurcation from the limit cycle to a fixed point by changing the parameter $\theta$.
From Eq. (\ref{JacobiMatrix}), the eigenvalues of the Jacobi matrix $\{\lambda\}$ are given by the solution of the following equation:
\begin{align}
    (-1 - \lambda)^3 + \frac{-\beta g}{\cosh^2(-\beta g (x_3 - \theta_1/g)}\cdot \frac{-\beta g}{\cosh^2(-\beta g (x_1 - \theta_2/g)} \cdot \frac{-\beta g}{\cosh^2(-\beta g (x_2 - \theta_3/g)} = 0. \label{determinant}
\end{align}
Here, we plotted $-\beta g/\cosh^2(-\beta g(x-z))$ as a function of $x$ in Fig. \ref{cosh}.
The $1/\cosh^2$ function has a tall spike at $x = z$ for a sufficiently large $\beta$.
Considering that we adopted the $\tanh$ function, sigmoidal function with range$\{-1, 1\}$, the value of $x_i$ is restricted to $\{-1, 1\}$.
This suggests that if even one of $\theta_i/g (i = 1, 2,3)$ in Eq. (\ref{determinant}) exceeds one, and the second term in Eq. (\ref{determinant}) is close to zero, and the solution to Eq. (\ref{determinant}) is only determined by $(-1 -\lambda)^3$, then $\lambda = -1$, corresponding to a stable fixed point.
Considering that Eq. (\ref{eigen_value_x}) gives the solution corresponding to an unstable fixed point, the bifurcation point from the limit cycle to the fixed point was determined as follows:
\begin{align}
    \max(\frac{\theta_1}{g}, \frac{\theta_2}{g}, \frac{\theta_3}{g}) = 1. \label{bifurcationpoint}
\end{align}

\subsection{Derivation of a bifurcation line in $\hat{\theta}_{u}-\hat{\theta}_{s_1}$ space \label{border}}
In Fig. 2b, a set of Hopf bifurcation points showed a straight line in the $\hat{\theta}_{u}-\hat{\theta}_{s_1}$ space.
Here, we estimated this line.
Each normalized eigenvector $\boldsymbol{v}_{u}, \boldsymbol{v}_{s_1}$ is given by $v_{u} = \{1, 1, 1\}/\sqrt{3}$ and $v_{s_1} = \{2, -1, -1\}/\sqrt{6}$.
Hence, the full model with fixed $\hat{\theta}_{u}, \hat{\theta}_{s_1}$ can be written as 
\begin{subequations}
\begin{align}
    \frac{d x_1}{d t} &= \tanh \{ -\beta g(x_3 - \frac{\hat{\theta}_{u}/\sqrt{3} + 2\hat{\theta}_{s_1}/\sqrt{6}}{g})\} - x_1(=X_1), \\
    \frac{d x_2}{d t} &= \tanh \{ -\beta g(x_1 - \frac{\hat{\theta}_{u}/\sqrt{3} - \hat{\theta}_{s_1}/\sqrt{6}}{g})\} - x_2(=X_2), \\
    \frac{d x_3}{d t} &= \tanh \{ -\beta g(x_2 - \frac{\hat{\theta}_{u}/\sqrt{3} - \hat{\theta}_{s_1}/\sqrt{6}}{g})\} - x_3(=X_3).
\end{align}
\end{subequations}

From Eq. (\ref{bifurcationpoint}), the condition of the bifurcation from the limit cycle to the fixed point is given by
\begin{widetext}
\begin{align}
    \max (\frac{\hat{\theta}_{u}/\sqrt{3} + 2\hat{\theta}_{s_1}/\sqrt{6}}{g}, \frac{\hat{\theta}_{u}/\sqrt{3} - \hat{\theta}_{s_1}/\sqrt{6}}{g}, \frac{\hat{\theta}_{u}/\sqrt{3} - \hat{\theta}_{s_1}/\sqrt{6}}{g}) = 1. \label{condition}
\end{align}
\end{widetext}
We considered a case in which $\hat{\theta}_{s_1} > 0$ in Fig. 2b, then Eq. (\ref{condition}) can be written as
\begin{align}
    \frac{\hat{\theta}_{u}/\sqrt{3} + 2\hat{\theta}_{s_1}/\sqrt{6}}{g} = 1.
\end{align}
Hence, the border line for $g = 0.4$ is given by
\begin{align}
    \hat{\theta}_{s_1} = -\frac{1}{\sqrt{2}} \hat{\theta}_{u} + \frac{\sqrt{6}}{5}.
\end{align}

\subsection{Derivation of stagnation point along $\boldsymbol{v}_{u}$\label{stagnation}}
\begin{figure}[htbp]
    \centering
    \subfigure%[]
    {
        \includegraphics[width=0.6\linewidth]{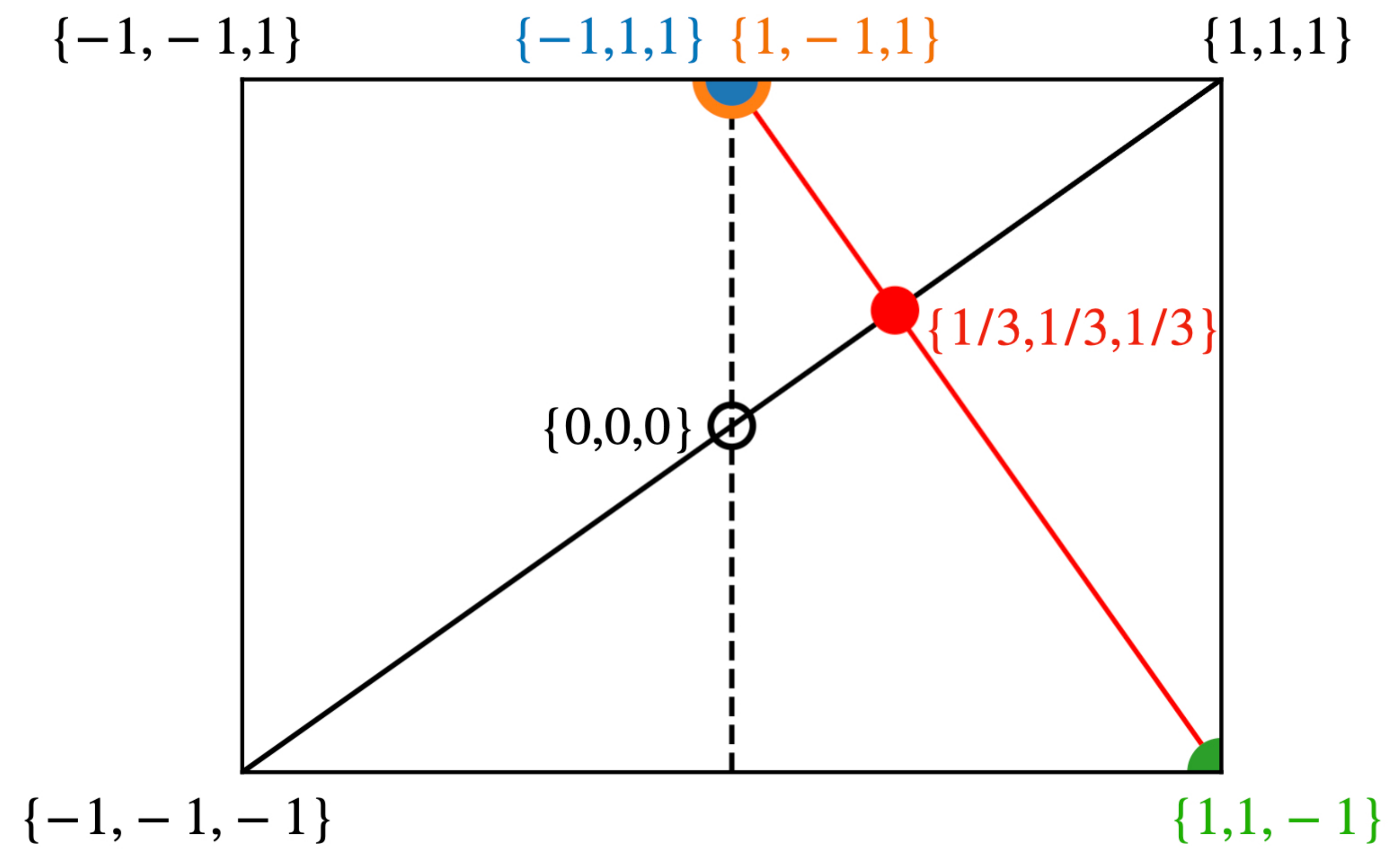}
    }
    \caption
    {
    Geometric representation for estimating stagnation point along $\hat{\theta}_{u}$.
    $x$ space was sliced on the $x_1 = x_2$ plane.
    }
    \label{geo}
\end{figure}

As shown in Fig. 2c, nonlinear slowing down occurred along the $\hat{\theta}_{u}$ axis.
In the repressilator model, we could estimate the stagnation point, which slowed down, that is, $d \hat{\theta}_{u}/d t \simeq 0$.
When $\hat{\theta}_{u}$ increased, the limit cycle orbit moved around differentiated fixed points $\{-1, 1, 1\}, \{1, -1, 1\}, \{1, 1, -1\}$ and was close to a triangular orbit with corners $\{-1, 1, 1\}, \{1, -1, 1\}, \{1, 1, -1\}$, as shown in Fig. 3a.
Thus, we approximated the limit cycle orbit to an equilateral triangle.
Fig. \ref{geo} schematically shows this orbit ($x_1 = x_2$ plane was adopted for plotting).

Then, from the symmetry, the time average of $x_i(t)$ should be located at the red point in Fig. \ref{geo}.
Considering geometry, the position of the red point is given by $\{x_1, x_2, x_3\} = \{1/3, 1/3, 1/3\}$.
In the geometric representation shown in Fig. \ref{geo}, the value of $\hat{\theta}_{u}$ agreed with the distance from the origin to the point on the line $x_1 = x_2 = x_3$.
Hence, $d \hat{\theta}_{u}/d t$ was minimized at $\hat{\theta}_{u} = 1/\sqrt{3} \sim 0.58$, 
well agreeing with the simulation, as shown in Fig. 3c.

Reprogramming dynamics required slowing down along $\boldsymbol{v}_{u}$.
According to our analysis, oscillation in $x$ was necessary for this process.
Thus, we imposed the parameter $g$, the strength of the negative feedback loop in the repressilator, the condition that the oscillation dynamics remained at the stagnation point $\{\theta_1, \theta_2, \theta_3\} = \{1/3, 1/3, 1/3\}$.
Then, the Hopf bifurcation should occur beyond the stagnation point with an increase in $\hat{\theta}_{u}$.
From Eq. (\ref{bifurcationpoint}), we obtained
\begin{align}
     g > \frac{1}{3}
\end{align}

\begin{figure}[htbp]
    \centering
    \subfigure%[]
    {
        \includegraphics[width=0.3\linewidth]{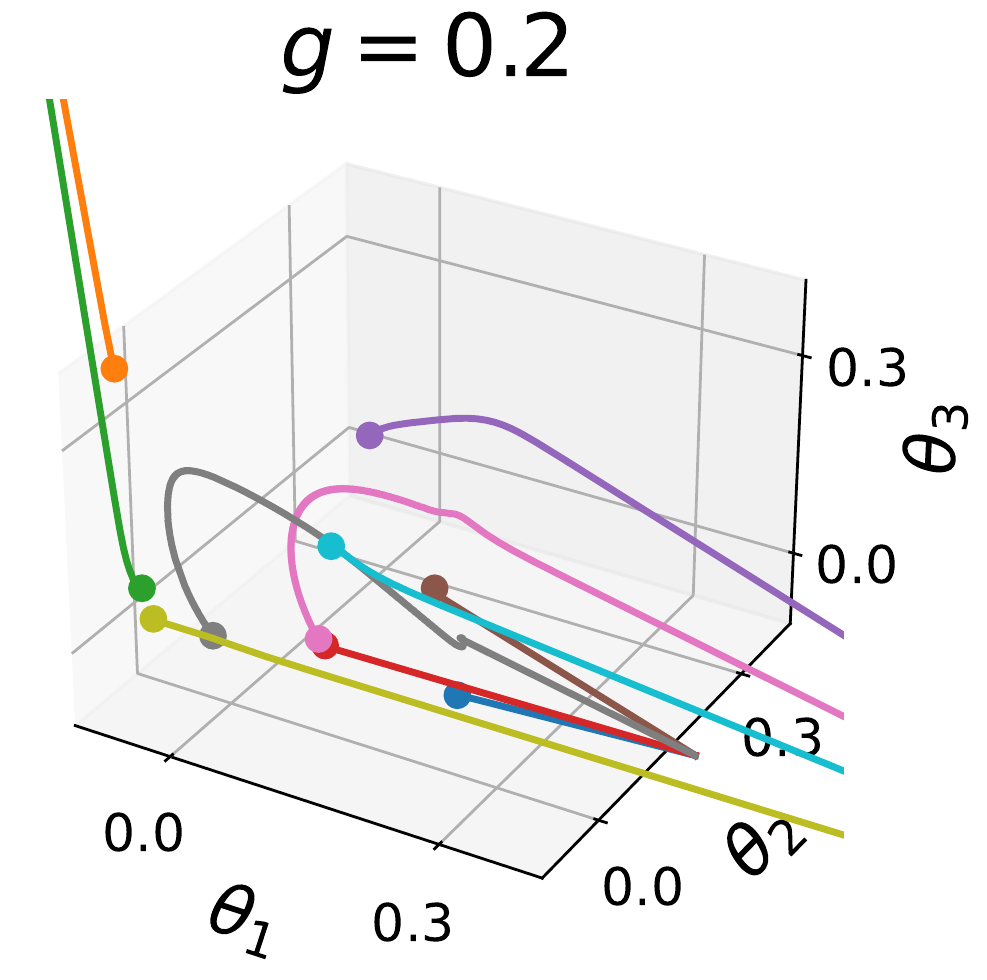}
        \includegraphics[width=0.3\linewidth]{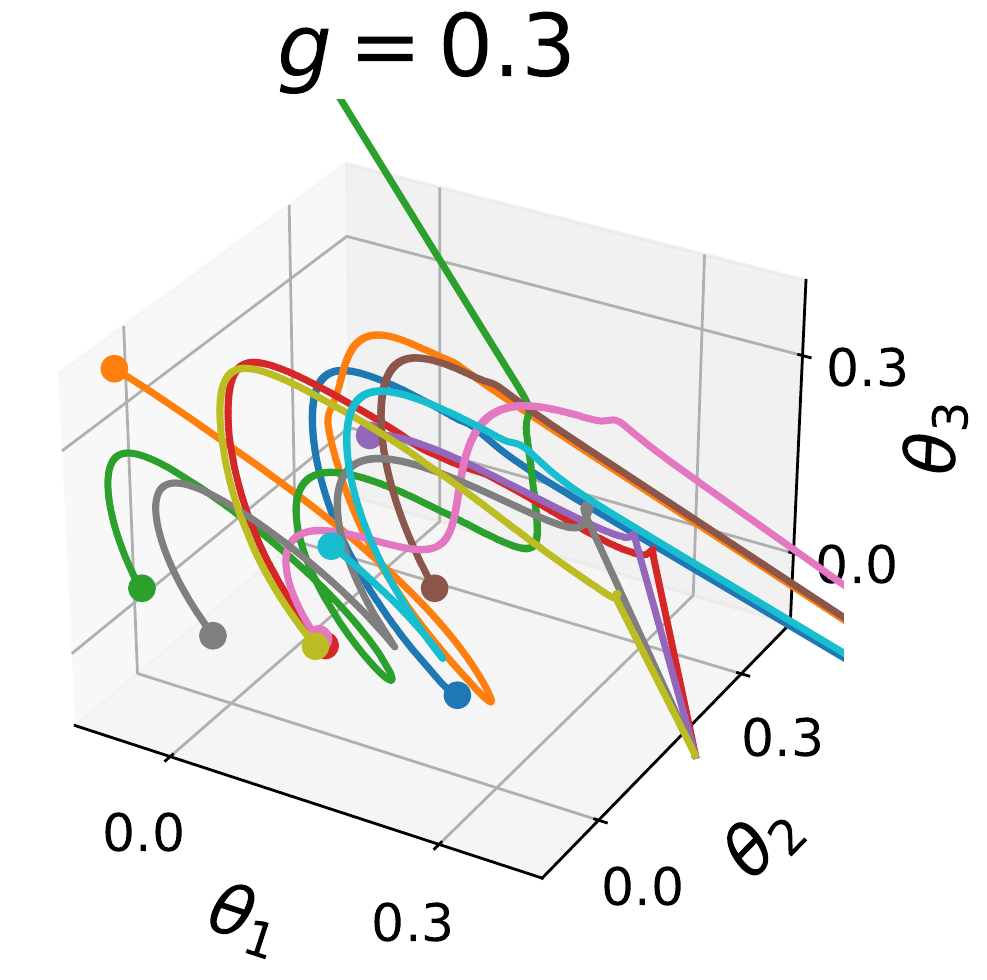}
        \includegraphics[width=0.3\linewidth]{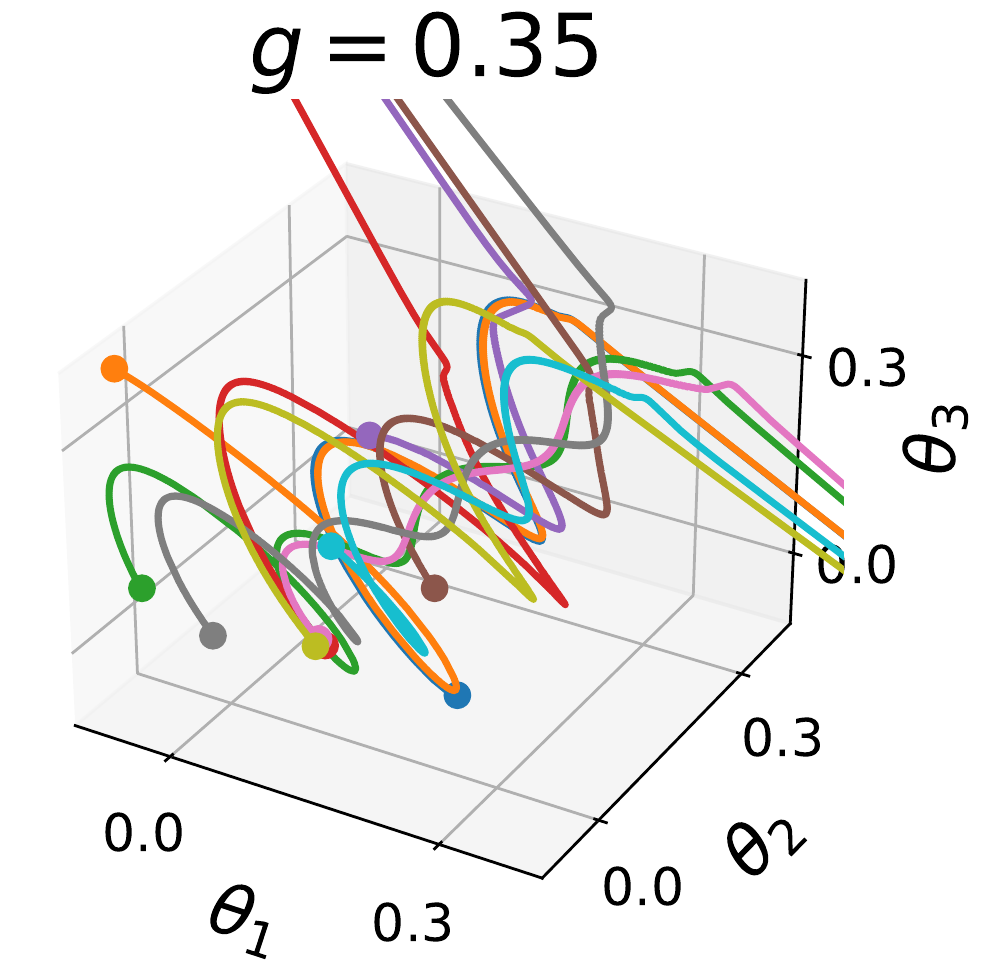}
    }
    \caption
    {
    Trajectories in the $\theta$ space, starting from the same initial conditions as in Fig. 1c, with $ g = 0.2, 0.3$, and $0.35$.
    }
    \label{g}
\end{figure}

\begin{figure*}[htbp]
    \centering
    \subfigure[]
    {
        \includegraphics[width=0.3\linewidth]{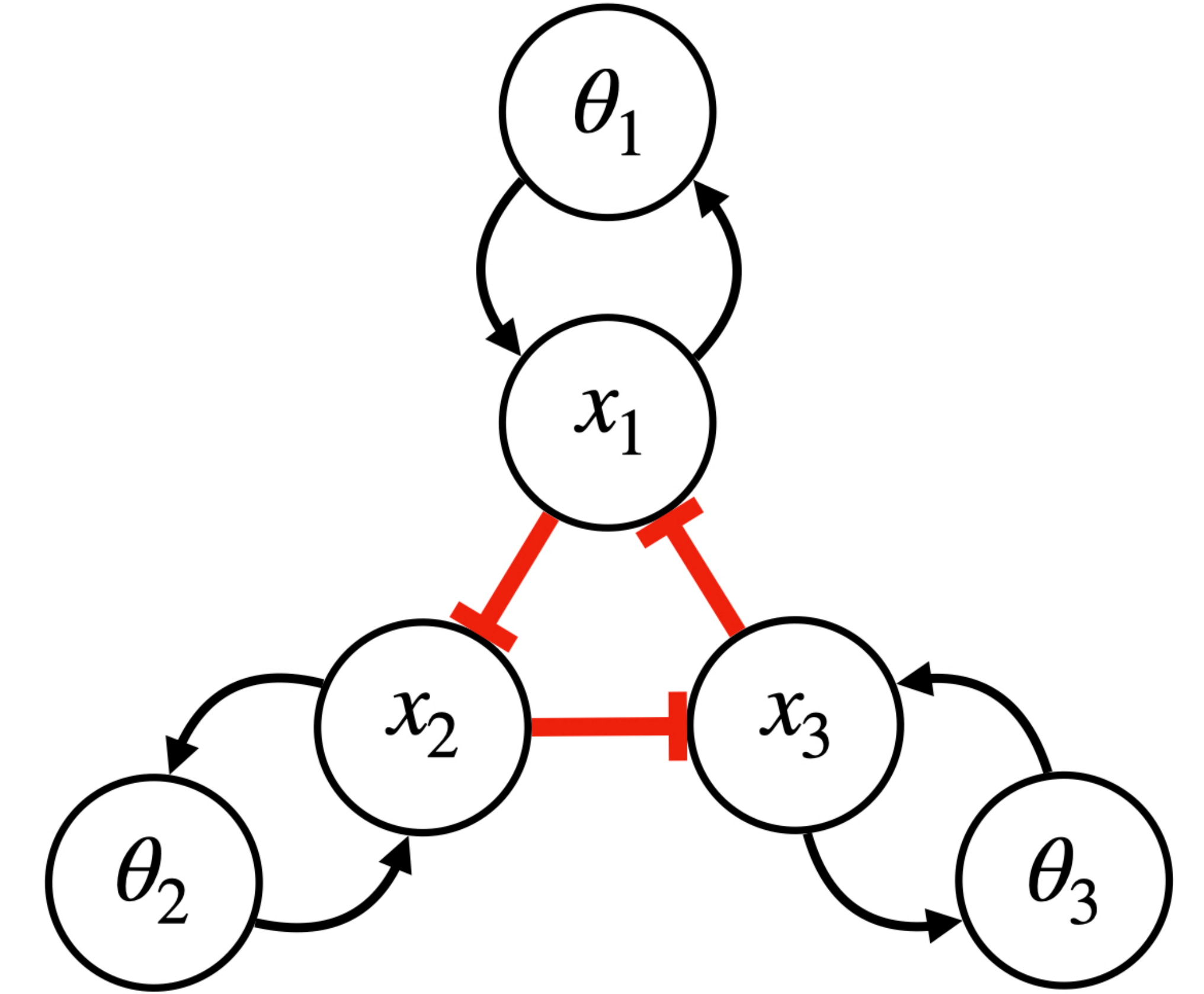}
    }
    \subfigure[]
    {
        \includegraphics[width=0.4\linewidth]{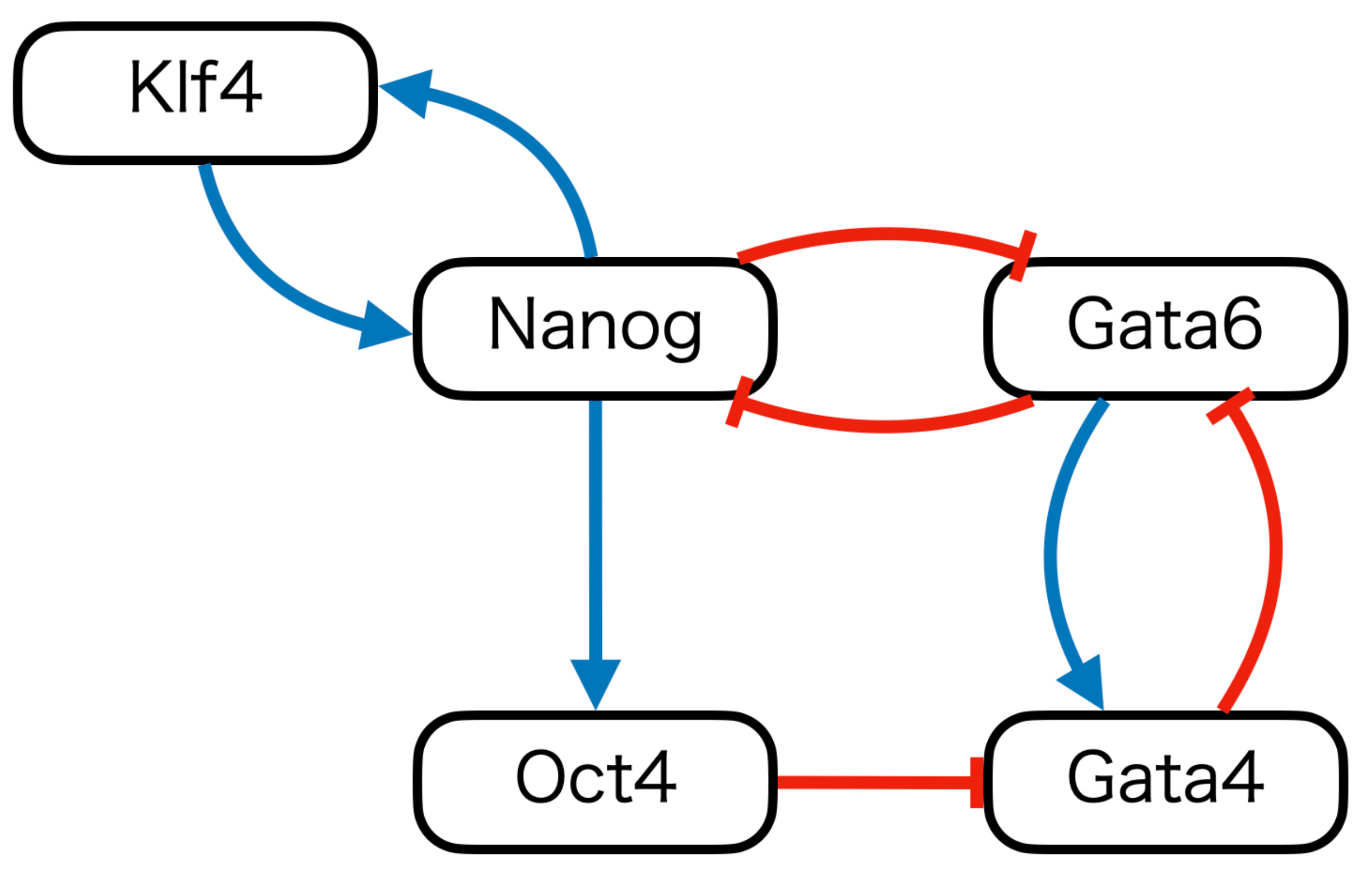}
    }
    \caption
    {
    Schematic representation of the cell model, which consisted of
$N$ genes, which state is represented by the gene expression $x_i$ and epigenetic modification status $\theta_i$.
    Red (blue) arrow shows the activation (inhibition) relation according to the gene regulatory matrix $J_{ij}$. 
    (a) The Repressilator Model comprised three genes that repress each other in a cyclic manner.
    $J_{21} = J_{32} = J_{13} = -g (=-0.4)$.
    Black arrow represents positive feedback relation between gene expression and epigenetic modification status.
    (b) Five-gene model. The $\theta_i$ nodes are not represented, but mutual positive feedback exists with the corresponding $x_i$, as in (a).
    \textit{Nanog, Oct4, Gata6, Gata4}, and \textit{Klf4} correspond to $x_1, x_2, x_3, x_4$, and $x_5$ with the following parameters: $J_{21} = 0.21, J_{42} = -1.1, J_{34} = -1.0, J_{43} = 0.46, J_{13} = -0.93, J_{31} = -0.16, J_{51} = 0.37, J_{15} = 0.64$, and the other $J_{ij} = 0$.
     }
    \label{Model}
\end{figure*}

As shown in Fig. \ref{g} and Fig. 1c, the global attraction occurred for $g = 0.35, 0.4$, whereas for $g = 0.2$ and $0.3$, $\theta$s went to fixed points without passing the stagnation point.
Then, reprogramming occurs for $g > 1/3$.

\newpage
\section{Supplemental figures}

\begin{figure*}[htbp]
    \centering
    \subfigure[]
    {
        \includegraphics[width=0.7\linewidth]{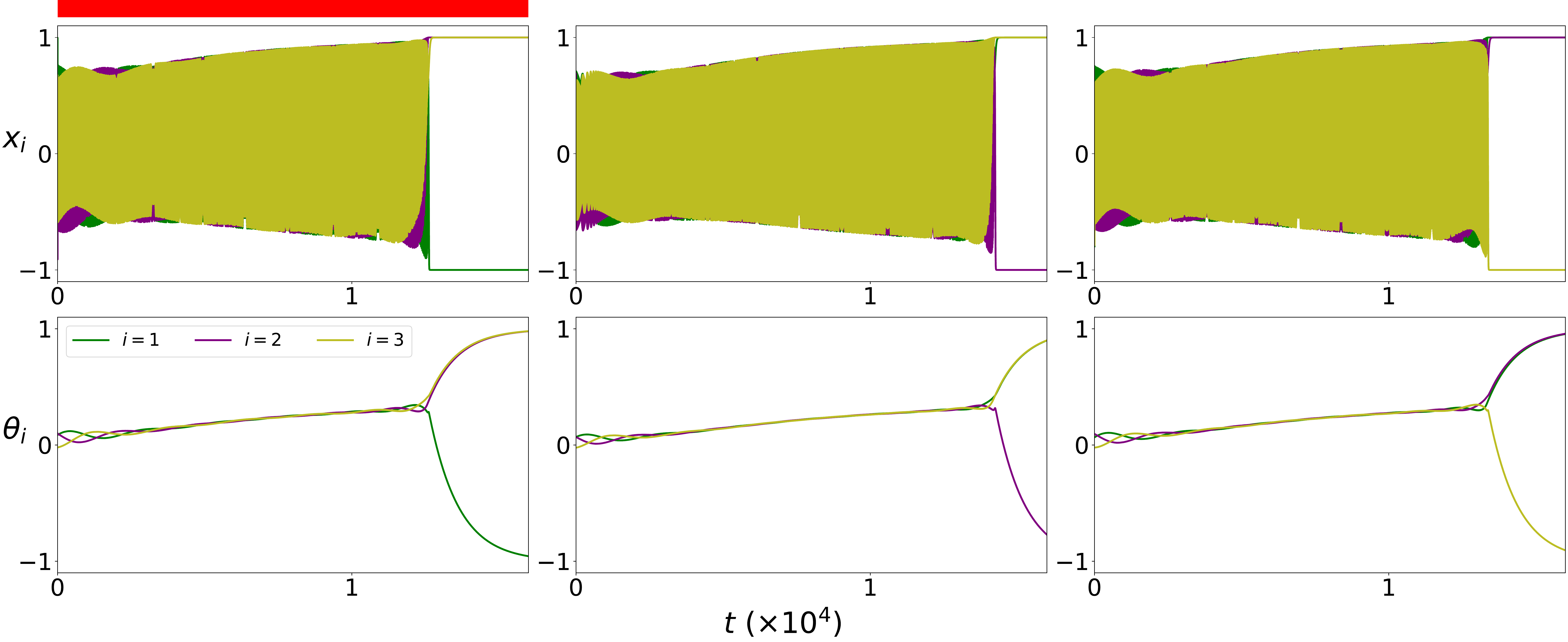}
    }
    \subfigure[]
    {
        \includegraphics[width=0.7\linewidth]{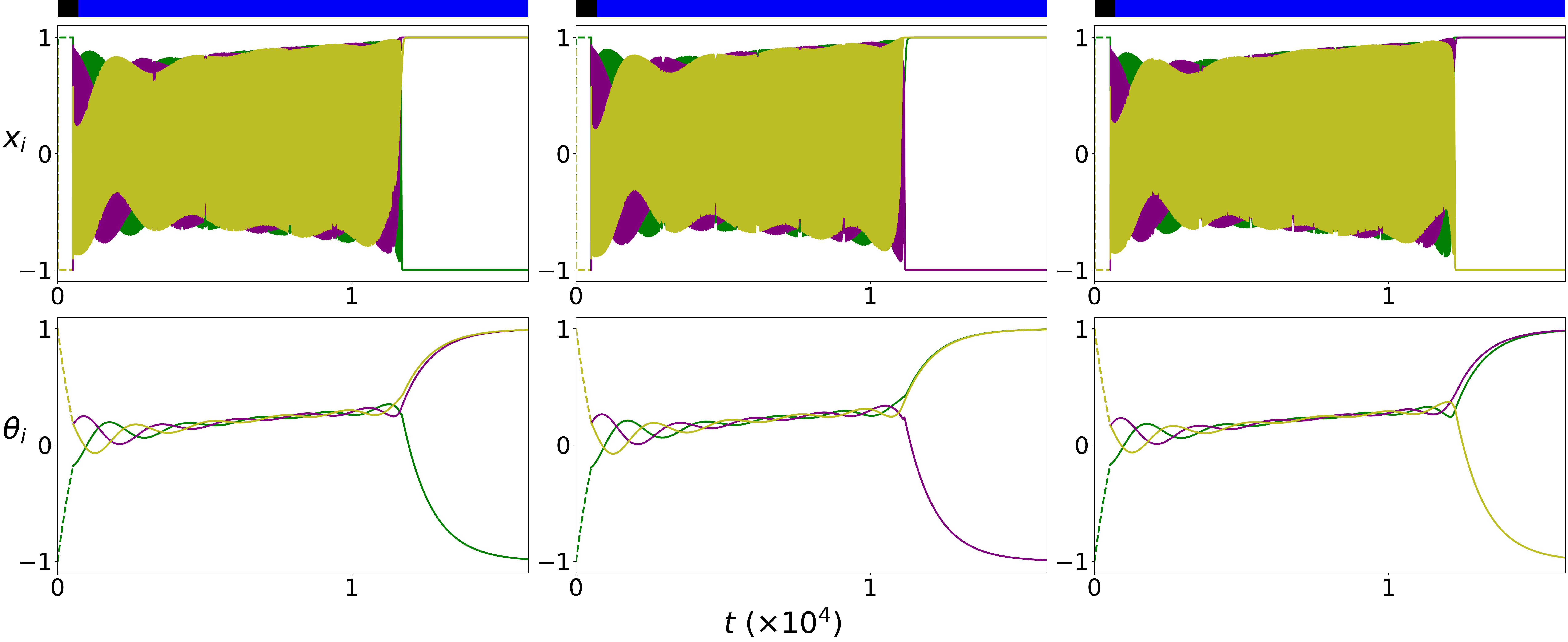}
    }
    \subfigure[]
    {
        \includegraphics[width=0.3\linewidth]{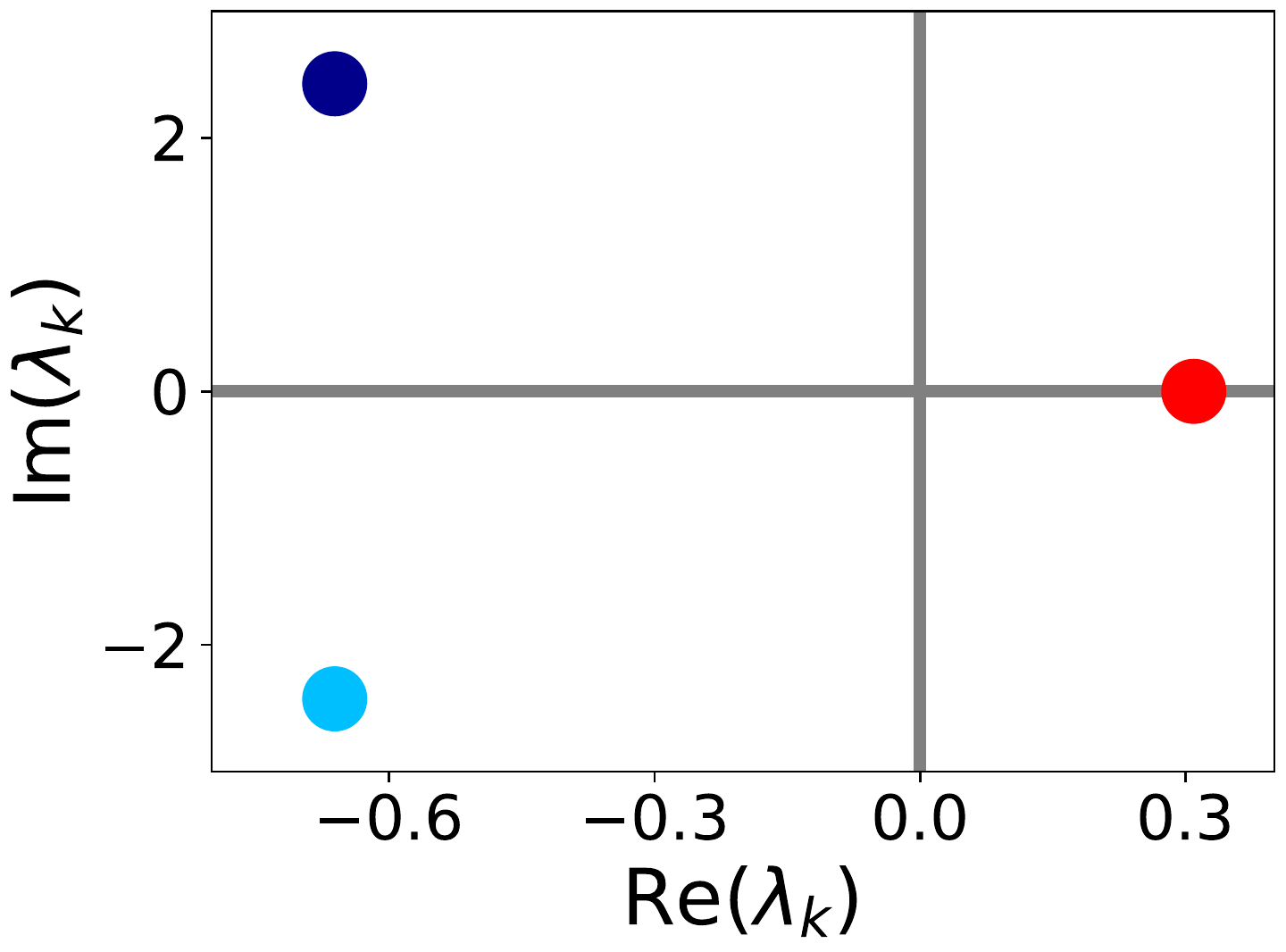}
    }
    \caption
    {
    Time development of all variables $x_i$ and $\theta_i$ of the (a) cell differentiation process and (b) reprogramming process in the repressilator model.
    (c) The eigenvalues $\lambda_k$ of the $\partial \Theta_i/\partial \theta_j$ matrix given by the fixed-point analysis of Eq. (2) around $\{\theta_1, \theta_2, \theta_3\} = \{0, 0, 0\}$.
    }
    \label{sup-rep}
\end{figure*}

\begin{figure*}[htbp]
    \centering
    \subfigure%[]
    {
        \includegraphics[width=0.7\linewidth]{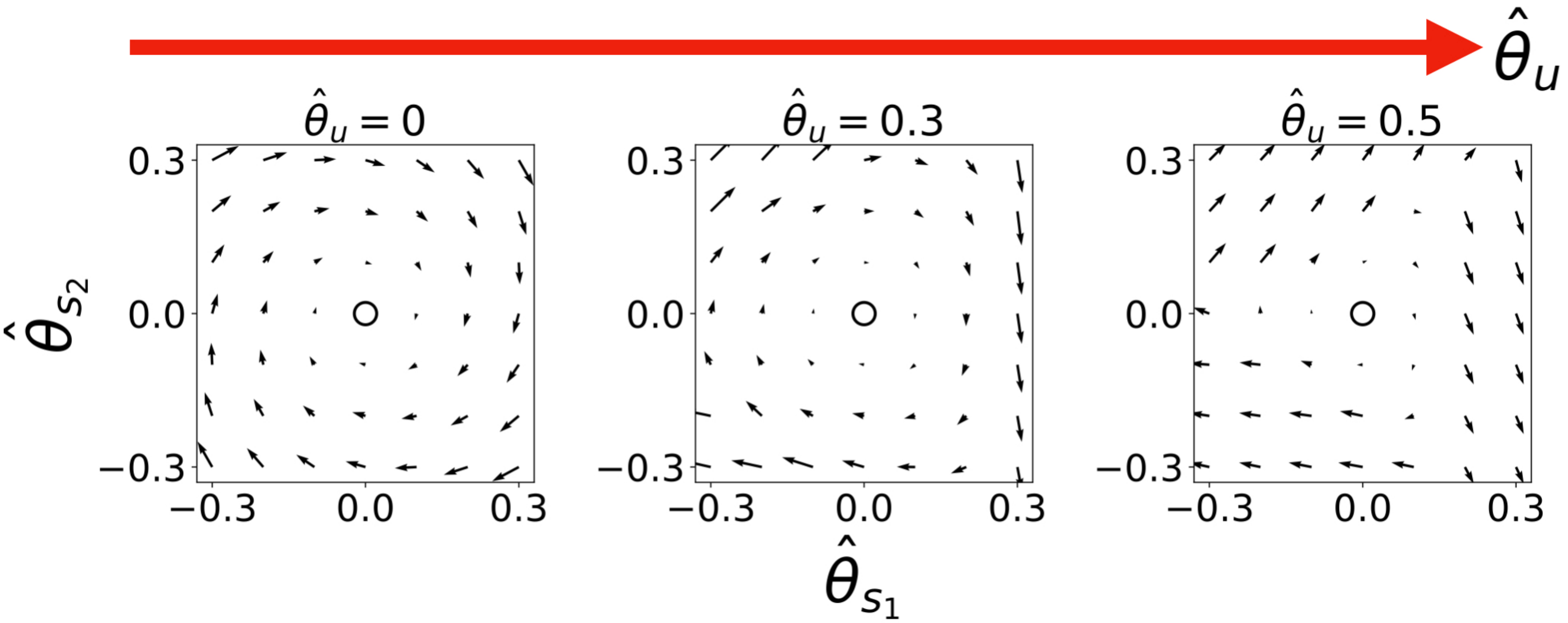}
    }
    \caption
    {
    The flow $(d \hat{\theta}_{s_1}/d t, d \hat{\theta}_{s_2}/d t)$ is plotted in the $(\hat{\theta}_{s_1}, \hat{\theta}_{s_2})$ plane along $\hat{\theta}_{u}$.
    Each center point $(0, 0)$ sits on the $\hat{\theta}_{u}$-axis.
    }
    \label{v12}
\end{figure*}

\begin{figure*}[htbp]
    \centering
    \subfigure[]
    {
        \includegraphics[width=0.5\linewidth]{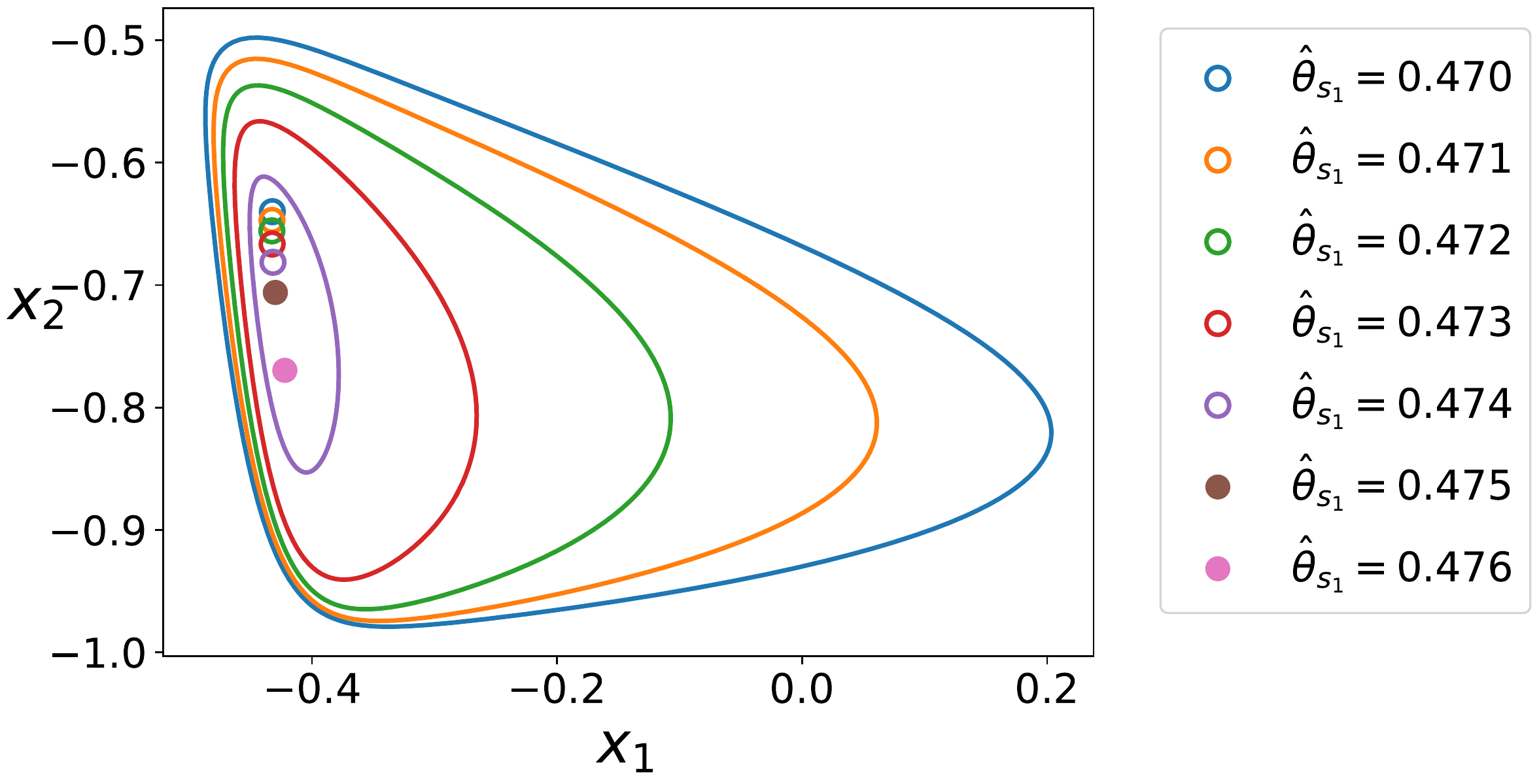}
    }
    \subfigure[]
    {
        \includegraphics[width=0.5\linewidth]{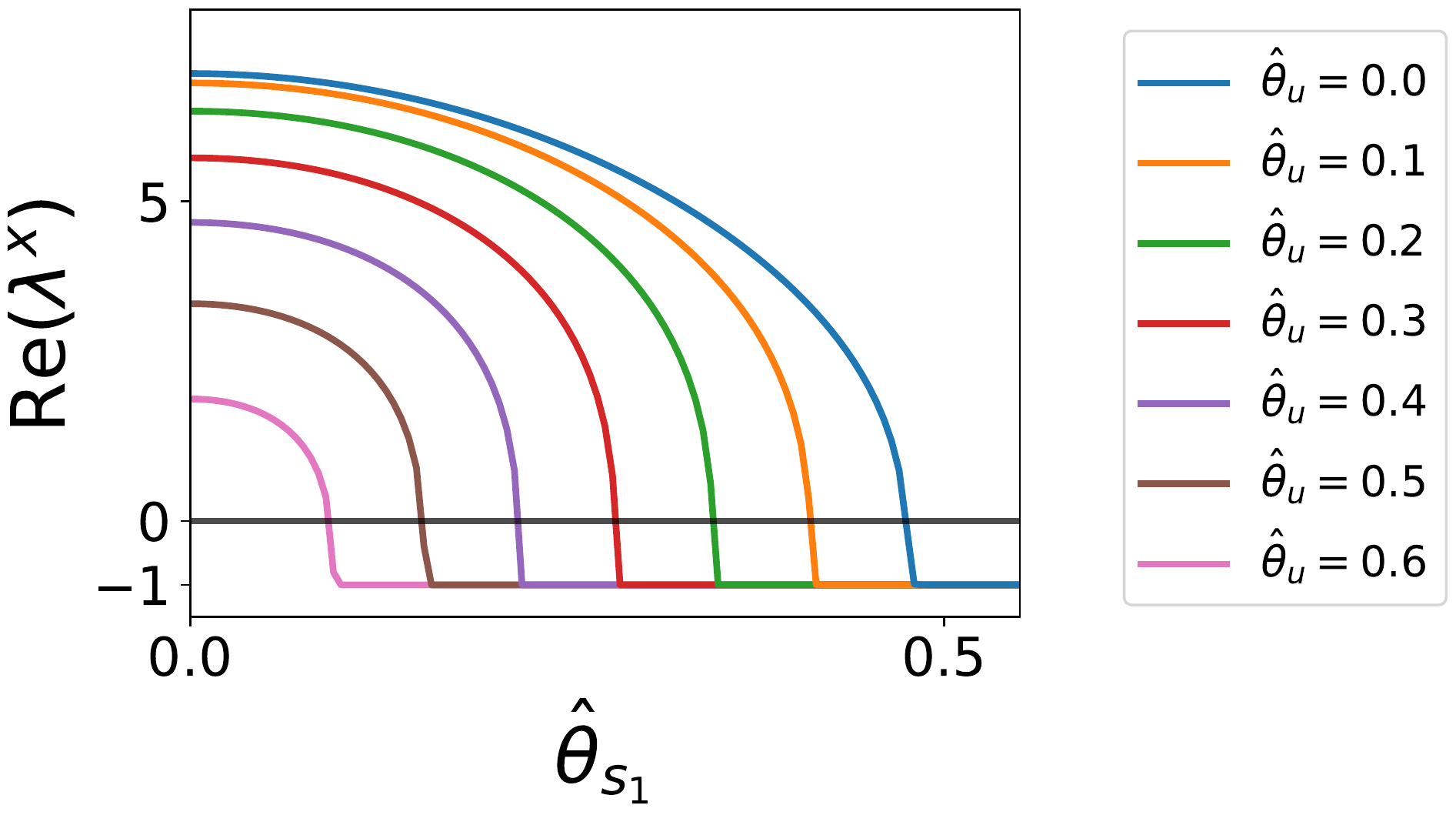}
    }
    \caption
    {
    (a) Change in the $x$ attractor in the vicinity of the bifurcation point against the change in $\hat{\theta}_{s_1}$ by fixing $\hat{\theta}_{u} = 0$.
    (b) Real part of eigenvalue in Eq. (1a) plotted as a function of $\hat{\theta}_{s_1}$ for various $\hat{\theta}_{u}$.
    Each line shows different value of $\hat{\theta}_{u}$.
    Bifurcation from the limit cycle to the fixed point occurs when the curve crosses the line $\mathrm{Re}(\lambda^x) = 0$.
    }
    \label{bifurcation}
\end{figure*}

\begin{figure*}[htbp]
    \centering
    \subfigure[]
    {
        \includegraphics[width=0.42\linewidth]{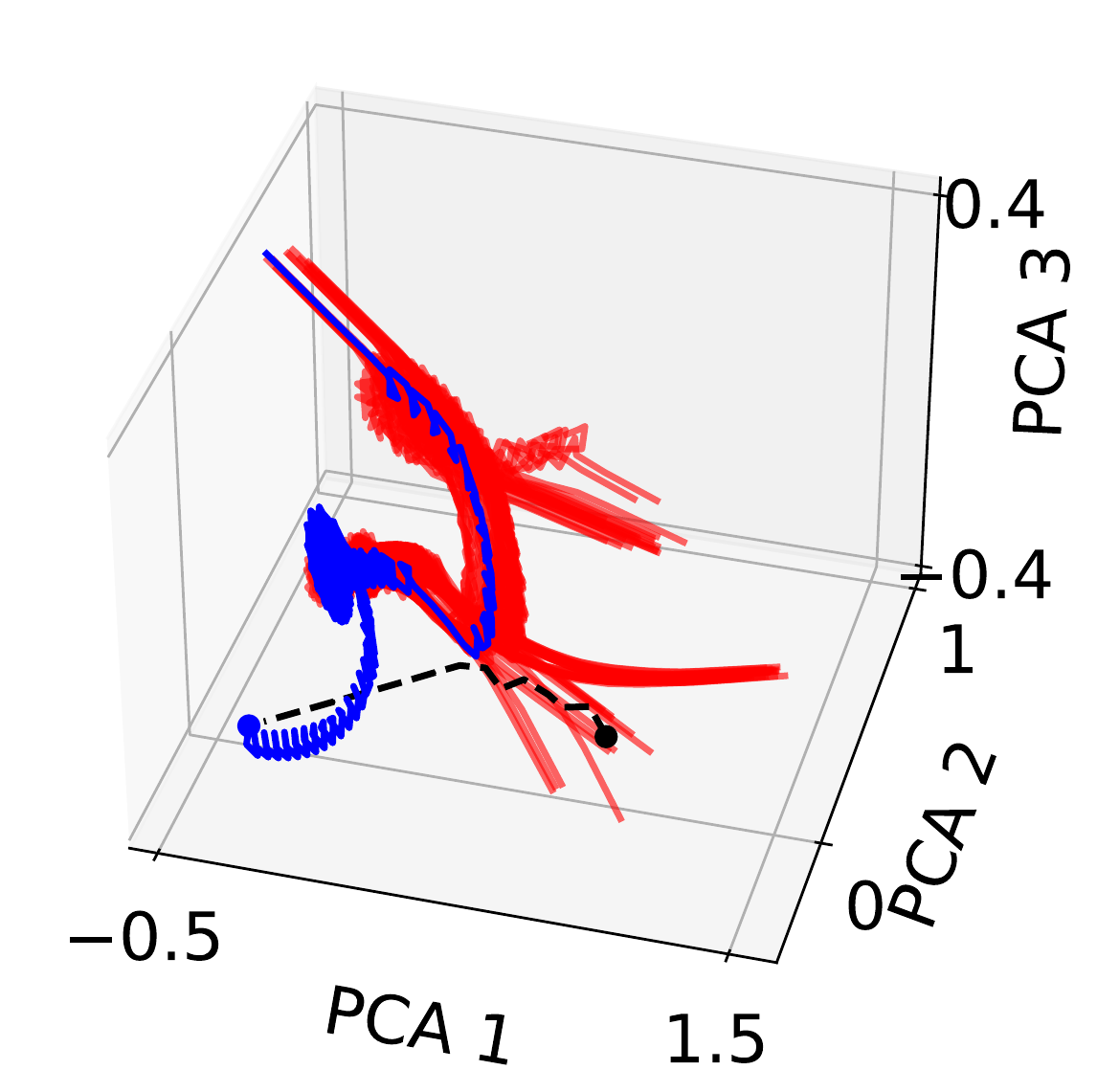}
    }\\
    \subfigure[]
    {
        \includegraphics[width=0.6\linewidth]{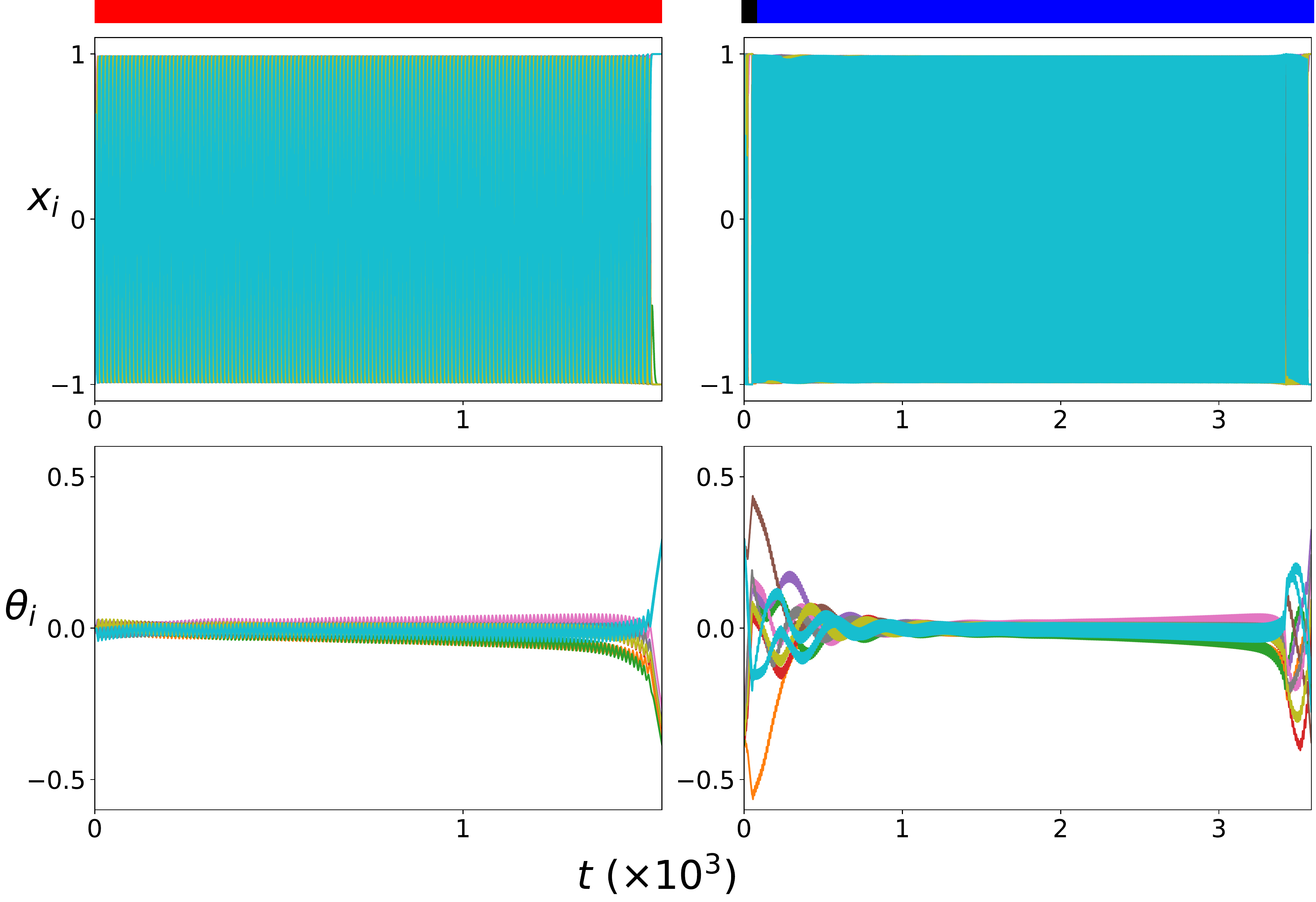}
    }
    \subfigure[]
    {
        \includegraphics[width=0.19\linewidth]{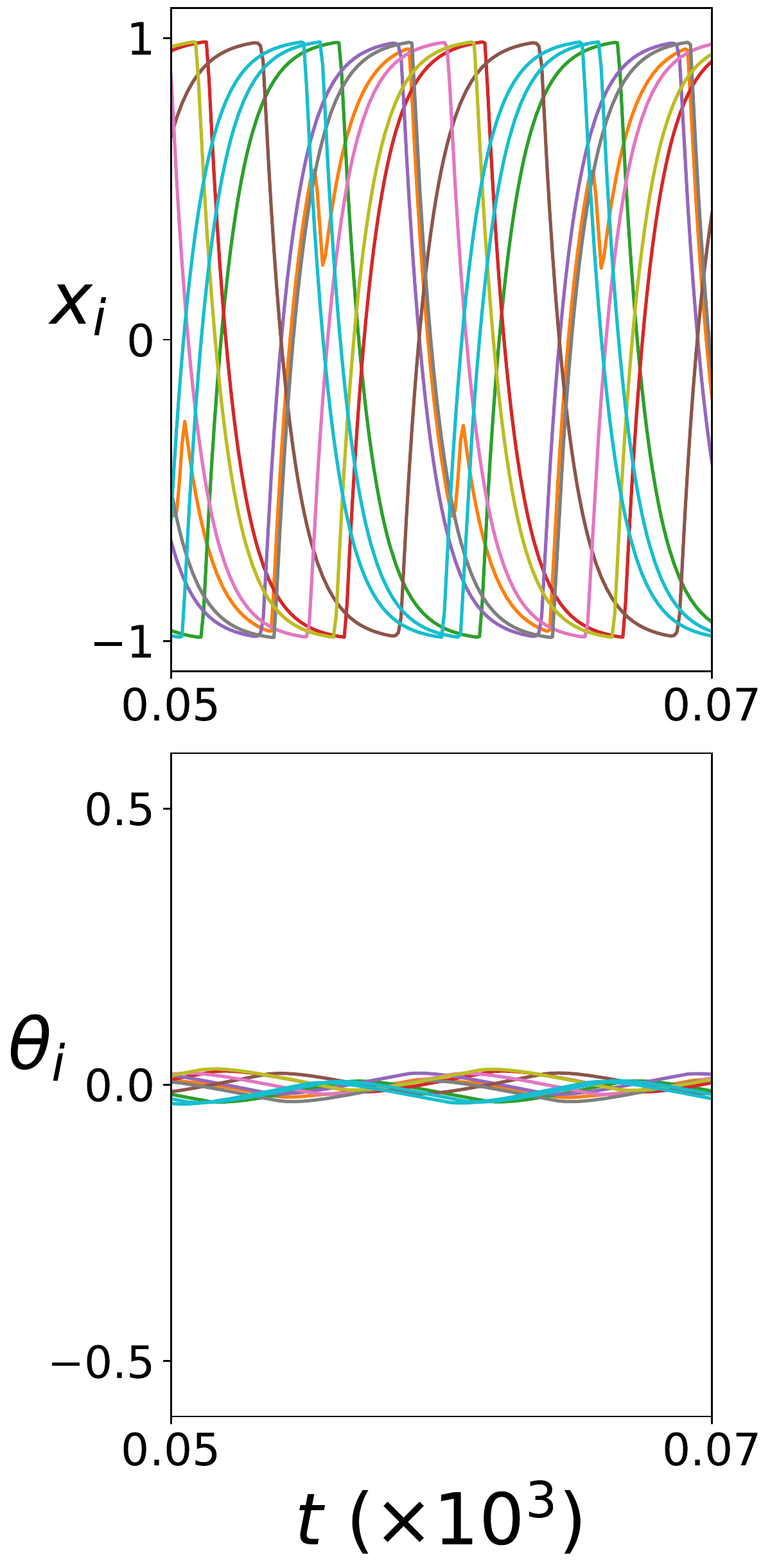}
    }
    \subfigure[]
    {
        \includegraphics[width=0.3\linewidth]{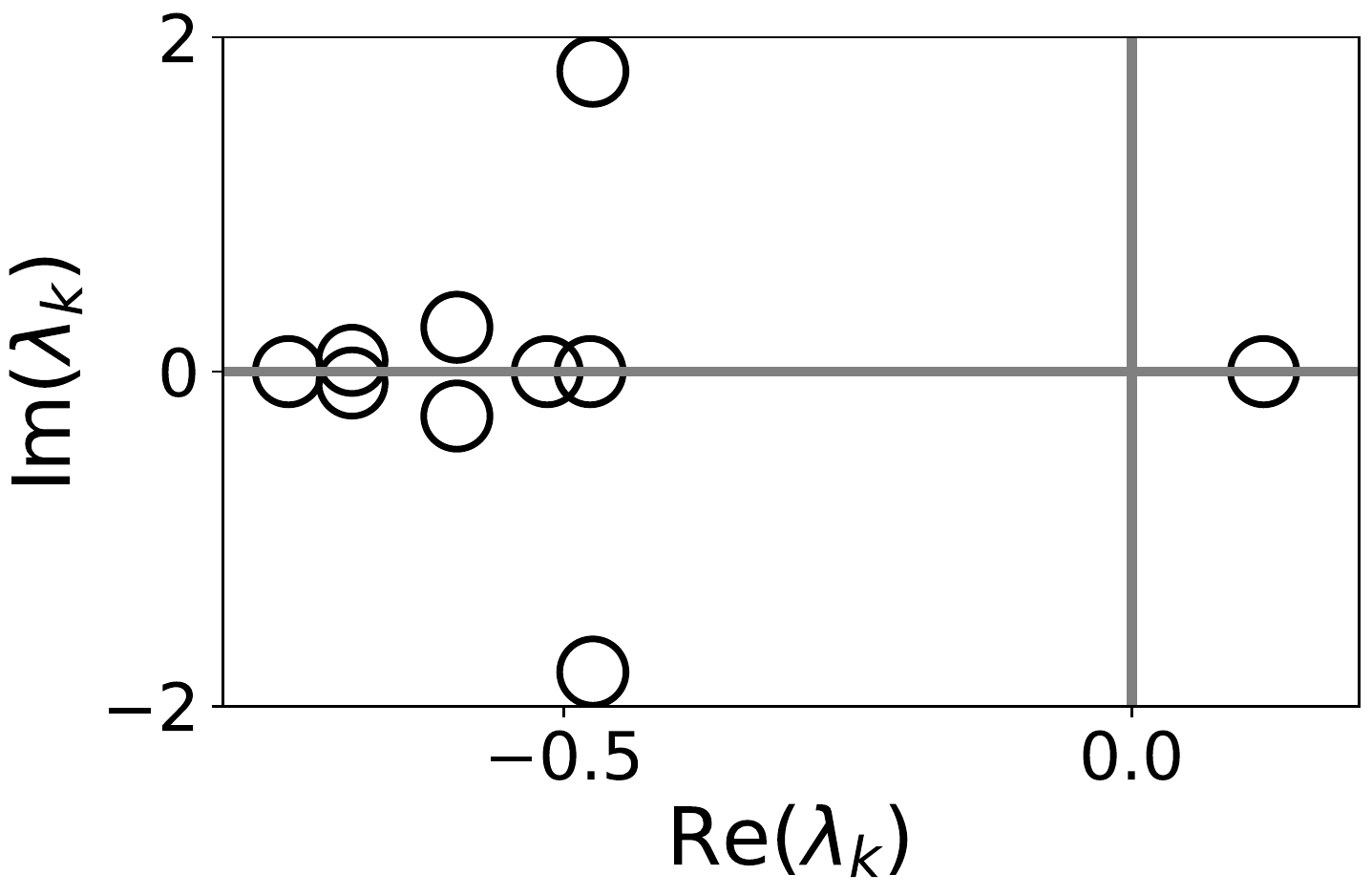}
    } \\
    \caption
    {
    Random gene regulatory network model with $N = 10$.
    We randomly generated the gene regulatory matrix $J_{ij}$ and selected one sample that showed oscillation dynamics for $x_i(t)$ around $\theta_i = 0$.
    $J_{ij}$ was generated according to the normal distribution, with an average of zero and a standard deviation of $1/\sqrt{N}$.
    (a) Cell differentiation and reprogramming.
    For plotting, we adopted a principal component analysis (PCA) of $\{\theta_i\} (i = 1, 2, \dots, 10)$ over 1,000 trajectories starting from $\theta_i \sim 0$.
    The trajectory $\{\theta_i(t)\}$ was plotted by adopting the first, second, and third PCA modes.
    (Red lines) Trajectories starting from $\theta_i \sim 0$, five different fixed points are reached after transient oscillation dynamics.
    (Black dotted line) From one of the differentiated cell types, we added external input $I_i(t)$ to three genes as reprogramming manipulation.
    (Blue line) After reprogramming manipulation, the cell state was attracted to the unstable manifold and differentiated again.
    (b) Time development of cell differentiation and reprogramming (Fig. \ref{10-gene}a.
    The upper bar shows the differentiation and reprogramming process, similar to Fig. 1ab.
    (c) Time development of cell differentiation (red region in Fig. S8b) for a short time window.
    (d) The eigenvalues $\lambda_k$ of $\partial \Theta_i/\partial \theta_j$ matrix given by fixed point analysis of Eq. (2) around $\theta_i = 0$ for all $i$.
    }
    \label{10-gene}
\end{figure*}

\begin{figure*}[htbp]
    \centering
    \subfigure[]
    {
        \includegraphics[width=0.9\linewidth]{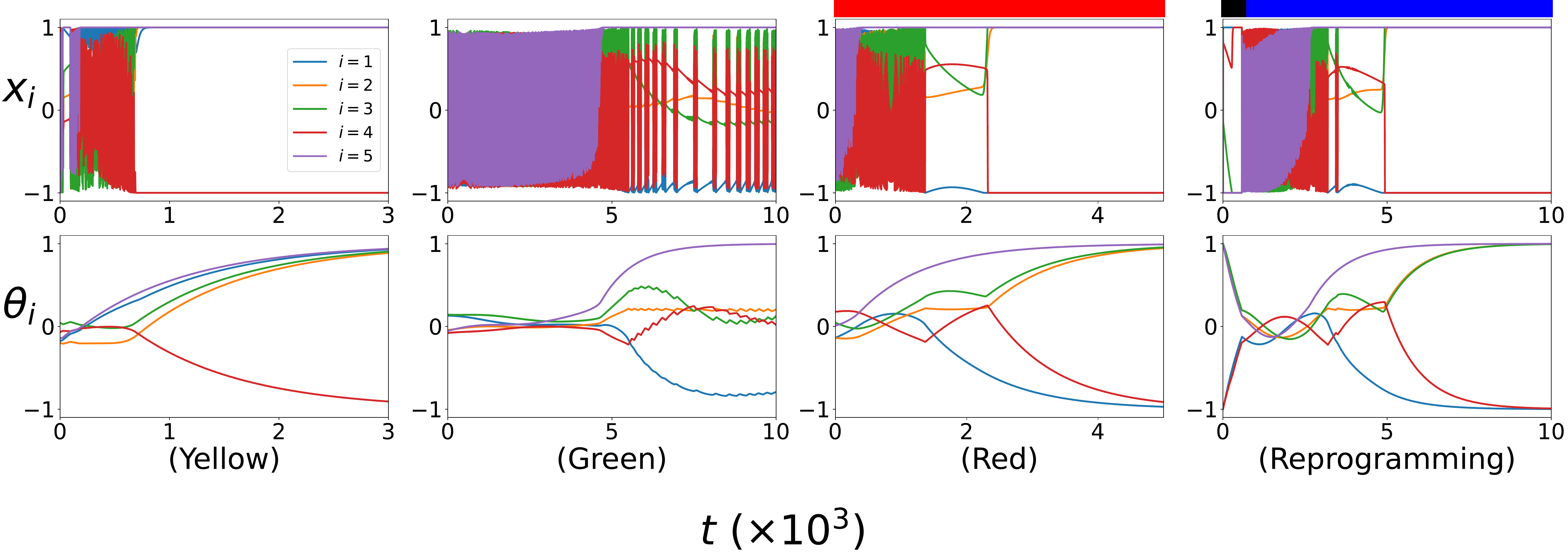}
    }
    \subfigure[]
    {
        \includegraphics[width=0.3\linewidth]{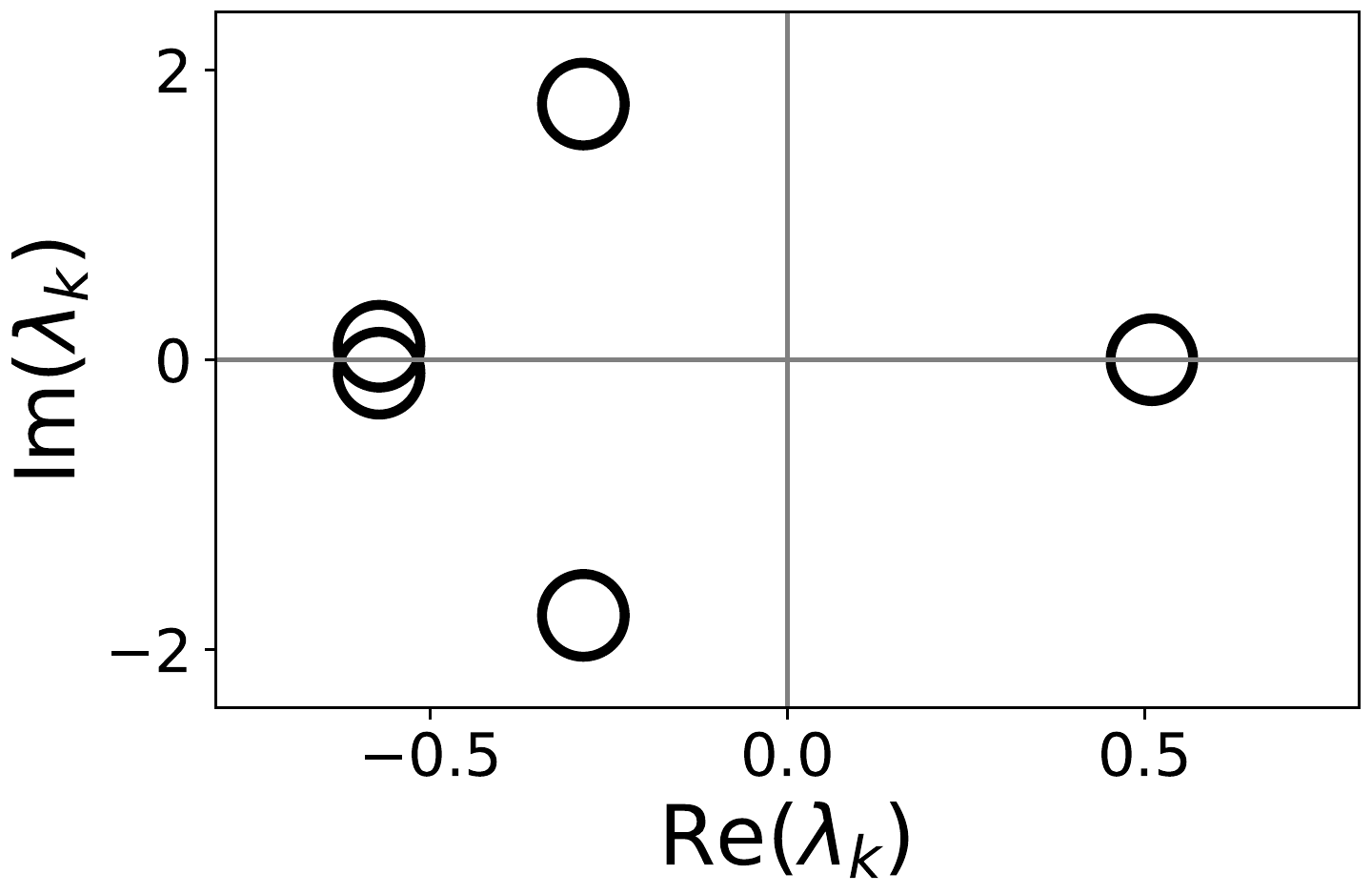}
    }
    \caption
    {
    (a) Time development for each trajectory of cell differentiation (yellow, green, and red line in Fig. 4a) and reprogramming (black dotted and blue lines in Fig. 4b).
    (Yellow, green, and red) Three trajectories starting from $\theta_i \sim 0$ are plotted, which reach three different attractors.
    (Black) From the differentiated state, we added an external input $I_i(t)$ to \textit{Nanog}($x_1$), \textit{Oct4}($x_2$), and \textit{Klf4}($x_5$) as reprogramming manipulation.
    (Blue) After reprogramming manipulation, the cellular state was first attracted to the unstable manifold, and then differentiated again.
    (b) The eigenvalues $\lambda_k$ of the $\partial \Theta_i/\partial \theta_j$ matrix given by the fixed-point analysis of Eq. (2) around $\theta_i = 0$ for all $i$.
    In Fig. 4, we adopted eigenvectors $\boldsymbol{v}_{s_1}, \boldsymbol{v}_{s_3}$ with different $\mathrm{Re}(\lambda_k)$.
    }
    \label{sup-5g}
\end{figure*}

% \nocite{*}
\bibliography{aipsamp}% Produces the bibliography via BibTeX.